\documentstyle[12pt,epsfig]{article}

\begin{document}
\newcommand {\be}{\begin{equation}}
\newcommand {\ee}{\end{equation}}
\newcommand {\ba}{\begin{eqnarray}}
\newcommand {\ea}{\end{eqnarray}}
\newcommand {\bea}{\begin{array}}
\newcommand {\cl}{\centerline}

\newcommand {\eea}{\end{array}}
\renewcommand {\thefootnote}{\fnsymbol{footnote}}

\vskip .5cm

\renewcommand {\thefootnote}{\fnsymbol{footnote}}
\def \a'{\alpha'}
\baselineskip 0.65 cm
\begin{flushright}
SLAC-PUB-9116\\
hep-ph/0201105 \\  
\today 
\end{flushright}
\begin{center}
{\Large{\bf Leptonic Unitarity Triangle and 
CP-violation}} 
{\vskip .5cm}
 Y. Farzan$^{1,2}$ and A. Yu. Smirnov$^{3,4}$

$^1$ {\it Scuola Internazionale Superiore di 
Studi Avanzati, SISSA, I-34014, Trieste, Italy}\\
$^2$ {\it Stanford Linear Accelerator Center, Stanford University, Menlo 
Park, California 94025}\\
$^3$ {\it The Abdus Salam International Centre for
Theoretical Physics, I-34100 Trieste, Italy}\\
$^4$ {\it Institute for Nuclear Research, RAS, Moscow, Russia}
\end{center}   

\begin{abstract} 
\baselineskip 0.5 cm
The area of the  unitarity triangle is a measure of  CP-violation. 
We introduce  the leptonic unitarity triangles and study 
their properties. 
We consider the possibility of reconstructing the
unitarity triangle in future oscillation and non-oscillation experiments. 
A set of measurements 
is suggested which  
will, in principle, allow us to measure all sides of the triangle, 
and consequently to establish  CP-violation. 
For different values of the CP-violating phase, $\delta_D$, 
the required accuracy of measurements   is estimated. 
The key elements of the method include determination of 
$|U_{e3}|$ and  studies  of 
the $\nu_{\mu} - \nu_{\mu}$ survival probability in  
oscillations driven by the solar mass splitting $\Delta m^2_{sun}$.  
We suggest additional astrophysical
measurements which may help to reconstruct the triangle. 
The method of the unitarity triangle is   complementary  to   
the direct measurements of  CP-asymmetry. It requires mainly studies of
the {\it survival} probabilities and processes where oscillations are averaged
or the coherence of the state is lost.   

\end{abstract} 
\baselineskip 0.65 cm
\section{Introduction} 

Measurement of CP-violation in leptonic sector is 
one of the main challenges in particle physics, 
astrophysics and cosmology. 

For three neutrinos (similarly to the quark sector \cite{qcp}) 
there is a unique complex phase in the lepton mixing matrix,  
$\delta_D$, which produces  observable CP-violating effects \cite{cpin}. 
(If neutrinos are Majorana particles, two additional CP-violating phases 
exist. These phases,   
the so-called Majorana phases,  do not appear in the
oscillation patterns.)  
The phase $\delta_D$ leads   to CP-asymmetry \cite{acp},  
$P(\nu_\alpha \rightarrow \nu_\beta) \neq P(\bar{\nu}_\alpha
\rightarrow \bar{\nu}_\beta)$, 
as well as T-asymmetry~\cite{at}, 
$P(\nu_\alpha \rightarrow \nu_\beta) \neq P(\nu_\beta \rightarrow \nu_\alpha)$, 
of the oscillation probabilities
(see  also \cite{akhmed} and references  therein).

Measurements of  the CP- and T- asymmetries provide a 
{\it direct} method of establishing   CP-violation. 
There are a number of studies of experimental 
possibilities to measure the asymmetries. 
It was realized that in the $3\nu$-schemes of neutrino mass and mixing 
which explain the atmospheric and solar neutrino data, the 
CP-violation and T-violation effects are small and 
it will be difficult to  detect them \cite{cpest}. 
The  smallness is due to 
small values of $U_{e3}$ (restricted by CHOOZ 
result) and $\Delta m_{sun}^2$ (responsible for 
the solar neutrino conversion). Still, the effect 
can be seen in the new generation of the long baseline 
(LBL) experiments provided that  the
LMA-MSW is the  solution of the 
solar neutrino problem and that $U_{e3} > 0.05$ 
\cite{golden,jhf-kamioka,hiba}.

Two  types of LBL experiments sensitive to $\delta_D$ 
are under consideration \cite{burton}: the experiments with superbeams 
\cite{jhf-kamioka,hiba} 
and neutrino beams from  muon storage rings (the neutrino factories) 
\cite{storage}.  
Analysis shows~\cite{jhf-kamioka,gomez} that  for $|U_{e3}|>0.05$ and 
$\Delta m_{sun}^2=5 \times 10^{-5} \ \ {\rm eV}^2$, neutrino factories can
discriminate between $\delta_D = 0$ and 
$\delta_D=\frac {\pi}{2}$ at the $3\sigma$  level \cite{whisnant}
while according to
\cite{jhf-kamioka} superbeams are able  to distinguish  at the $3\sigma$
level  $\delta_D=0$ from $\delta_D=\frac {\pi} {9}$.  
In  these experiments, the sensitivity to $\delta_D$  
decreases linearly with $\Delta m_{sun}^2$. So, the present uncertainty in
$\Delta m_{sun}^2$ results in an order of magnitude uncertainty in
evaluation of
sensitivity to $\delta_D$ in the future neutrino factories and superbeam
experiments. 
If $\Delta m_{sun}^2$ is smaller than  $2 \times 10^{-5}$ eV$^2$, the
direct methods will not be sensitive to $\delta_D$ \cite{golden}.  
Moreover,  neutrino factories and superbeams are very expensive and
technically difficult,  interpretation of  their results 
can be rather complicated and  ambiguous. In view of these 
difficulties, we need  to explore any alternative 
way to search for  CP-violation.  

Notice that apart from the asymmetries, the phase $\delta_D$ 
can be determined also 
from measurements of CP-conserving quantities,  
the oscillation probabilities themselves, which depend on 
$\delta_D$ \cite{lipari}.

The alternative method  to establish  CP-violation is to measure
the area  of  {\it unitarity triangle}. 
This method is well elaborated  in the quark sector.  
Indeed, the area of the unitarity triangle, $S$,  is related to  the
Jarlskog invariant, $J_{CP}$, which is a parameterization independent
measure of  CP-violation, as 
\be
S = \frac{1}{2} J_{CP}. 
\label{jcp}
\ee  
So, to establish CP-violation 
it is sufficient to show that the longest side of the triangle,  
is smaller than the sum of the other two.  

The problem is  to  measure   lengths of 
the sides of the triangle. As we will see, 
the method of the unitarity triangle  differs from 
measurements of asymmetries and may have certain advantages  
from the experimental point of view.

Previously, 
some general  properties of  the unitarity triangles for
leptonic sector (geometric features, test of unitarity)
have  been discused in \cite{sato}, \cite{xing}, \cite{branco}.


In this paper we will consider the possibility to reconstruct the leptonic
unitarity triangle. In sect.~2, we introduce the leptonic unitarity
triangles 
and study their properties. We estimate the accuracy with which 
the sides of the triangle  should be measured to establish 
CP-violation. In sect.~3, we describe a set of oscillation measurements
which would in principle allow us to reconstruct the triangle. 
  Additional astrophysical  measurements 
which would allow us to realize the method are suggested in sect.~4. 
Discussions and conclusions are given in sect.~5.

\section{Leptonic unitarity triangles}

In the three-neutrino schemes  
the flavor neutrino states, $\nu_f \equiv (\nu_e, \nu_{\mu}, \nu_{\tau})$, 
and the mass eigenstates $\nu_{mass} \equiv (\nu_1, \nu_2, \nu_3)$, 
are related by the unitary MNS (Maki-Nakagawa-Sakata \cite{mns}) 
matrix~\footnote{The mixing of three flavor states (two light neutrinos 
and heavy neutral lepton from the third generation) have been discussed in 
\cite{shrock}.}:
\ba
\label{mat}
U_{MNS} =   \left[ \matrix{ U_{e1} & U_{e2} & U_{e3} \cr
 U_{\mu 1} & U_{\mu 2} & U_{\mu 3} \cr
 U_{\tau 1} & U_{\tau 2} & U_{\tau 3} \cr}
\right].
\ea
The unitarity implies 
\ba
\matrix{U_{e1}U_{\mu 1}^* +U_{e2} U_{\mu 2}^*+U_{e3}U_{\mu 3}^*=0,  \cr
U_{e1}U_{\tau 1}^* +U_{e2} U_{\tau 2}^*+U_{e3}U_{\tau 3}^*=0,  \cr 
U_{\tau 1}U_{\mu 1}^* +U_{\tau 2} U_{\mu 2}^*+U_{\tau 3}U_{\mu 3}^*=0. 
}
\label{unit}
\ea 
In the complex plane, each term from the sums  
in (\ref{unit}) determines a vector. So,  the Eqs. (\ref{unit})  
correspond to  three unitarity  triangles. 
The  CP-violating phase, $\delta_D$,  vanishes 
if and only if   phases of all elements of 
matrix (\ref{mat}) are factorizable: 
$U_{\alpha i}=e^{i(\sigma_\alpha +\gamma_i)}|U_{\alpha i}|$. In this 
case $ U_{\alpha i}U_{\beta i}^*=e^{i( 
\sigma_\alpha-\sigma_\beta)}|U_{\alpha i}||U_{\beta i}|$,  
and therefore the unitarity triangles shrink to segments.

To construct the  unitarity triangle,  one  needs  
to  measure the absolute values of the elements of 
two rows (or equivalently two columns)   
in the mixing matrix. 
The area of the triangle is given by the Jarlskog invariant, $J_{CP}$
Eq. (\ref{jcp}). The area  is non-zero only if $\sin\delta_D \neq 0$.

\subsection{$e- \mu$ triangle; properties}

We will consider the triangle formed by the 
$e$- and $\mu$-rows of  the matrix (\ref{mat}) 
(see Eq. (\ref{unit}-a)). 
(Up to now, there is no  direct information 
about the elements of the third row. 
Moreover,  even in future, 
both creation of intense $\nu_{\tau}$ beams 
and detection of $\nu_{\tau}$ seem to be  difficult.)   

To reconstruct the $e- \mu$ triangle three quantities should be  
determined independently:  
\be
|U_{e1} U_{\mu 1}^*|, ~~~~~ |U_{e2} U_{\mu 2}^*|,  ~~~~~|U_{e3} U_{\mu
3}^*|.   
\label{sides}
\ee

The form of the triangle depends on the yet unknown  value of 
$|U_{e3}|$  and on the specific solution of the solar 
neutrino problem.  In what follows, we will  consider  mainly the  
LMA-MSW solution which provides the best fit for the solar neutrino
data.

\begin{figure}[ht]
\centering\leavevmode
\epsfxsize=0.7\hsize
\epsfbox{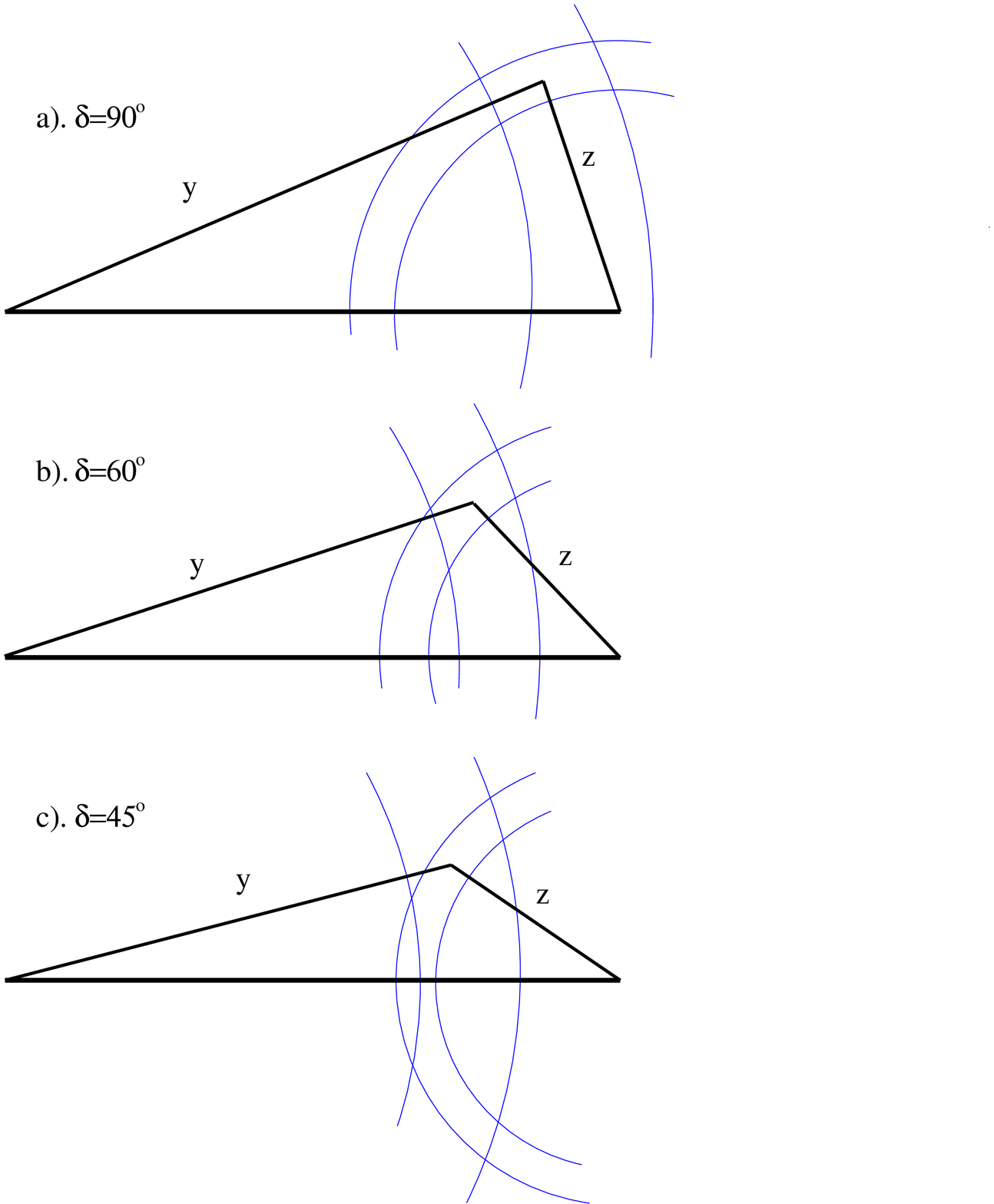}
\baselineskip 0.4 cm

\caption{
The unitarity triangles for different values of the 
CP-violating phase $\delta_D$. For mixing 
angles, we take $\tan^2 \theta_{12} = 0.3$, 
$\sin^2 2\theta_{23} = 1$ and $\sin^2 2\theta_{13} = 0.12$. 
The arcs  show the 10 \% uncertainties in determination of 
$y$ and $z$. 
}
\label{tri1}
\end{figure}
\baselineskip 0.65 cm

\begin{figure}[ht]
\centering\leavevmode
\epsfxsize=0.6\hsize
\epsfbox{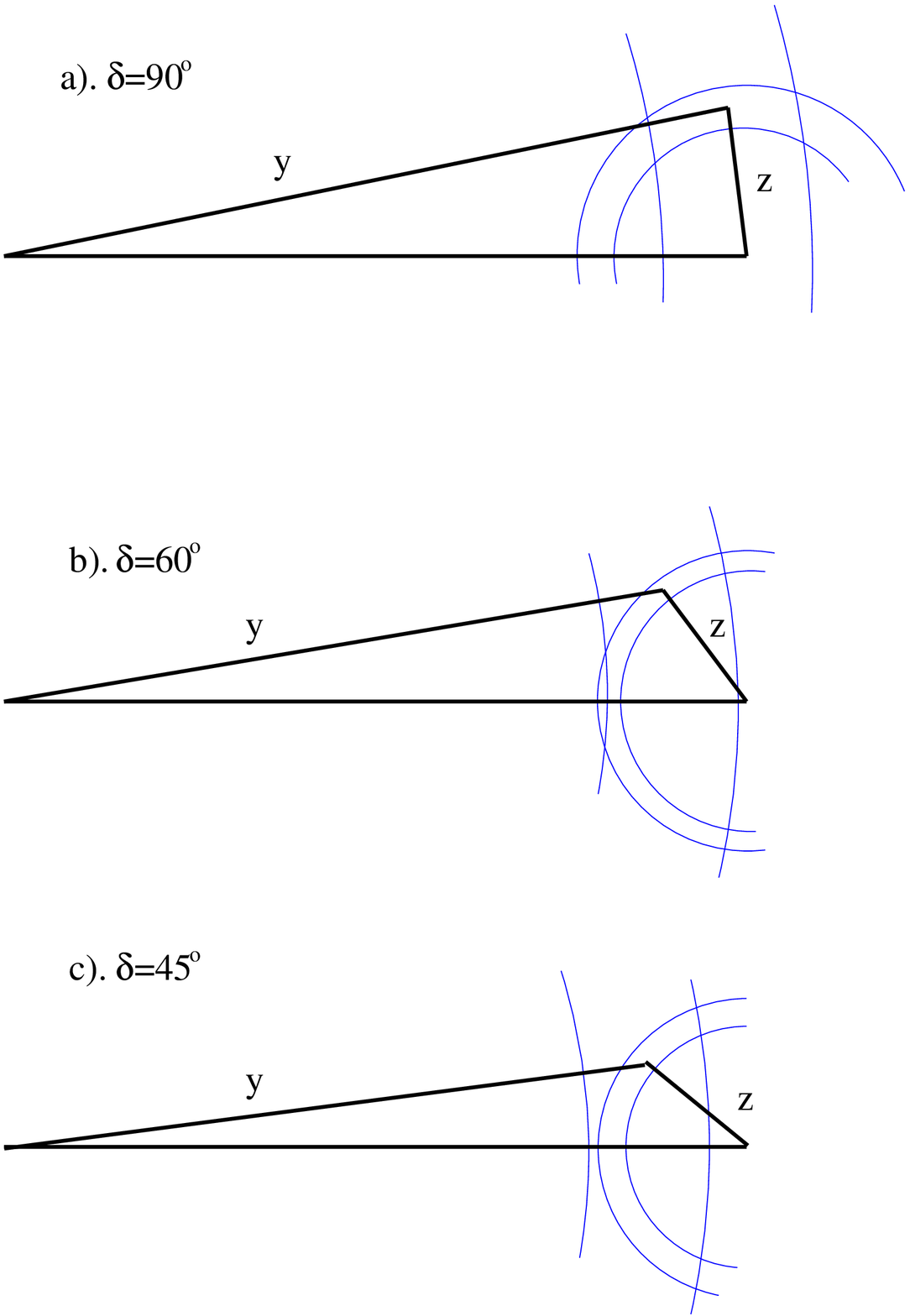}
\caption{
The same as  Fig. 1 for $\sin^2 2\theta_{13} = 0.03$.
}
\label{tri2}
\end{figure}

In Figs. ~\ref{tri1}  and  ~\ref{tri2}, we show examples of the unitarity
triangles  for different values of $U_{e3}$ and $\delta_D$.
In these figures we have normalized the sides of the triangles   in such a way 
that the length of the first side  equals one: 
\be
x = 1, ~~~ y = \frac{|U_{e2} U_{\mu 2}^*|}{|U_{e1} U_{\mu 1}^*|} ~~~ 
{\rm and} 
~~~
z = \frac{|U_{e3} U_{\mu 3}^*|}{ |U_{e1} U_{\mu 1}^*|}~. 
\label{xyz}
\ee
We  use  the standard parameterization of the 
MNS mixing matrix \cite{standard} in terms of the 
rotation angles $\theta_{12}$, $\theta_{13}$, $\theta_{23}$ 
and the phase $\delta_D$. We take  values of  $\theta_{12}$ and 
$\theta_{23}$  from the regions 
allowed  by  the solar and atmospheric neutrino  data.  

In Fig.~\ref{tri1} we present the triangles which correspond to 
$\sin^2  2\theta_{13} = 0.12$ (the upper
bound from the CHOOZ experiment for $\Delta m_{atm}^2=3\times 10^{-3} ~~
{\rm eV}^2$). The  arcs show  
10\% uncertainty in measurements of the sides $y$ and $z$. 
{}From Fig.~\ref{tri1}, one can conclude that 
for maximal CP-violation, $\delta_D = 90^{\circ}$, 
the existence of  CP-violation can be established at the $3\sigma$-level 
or even better if  the sides of the triangle are measured with 10\%
accuracy.  For $\delta_D = 60^{\circ}$,  the confidence level 
is approximately $2\sigma$. No statement can be made for 
$\delta_D \leq 45^{\circ}$
unless the accuracy of measurements of the sides will be better.   
These estimates should be considered as tentative ones.  
In order to make precise statements one needs to  perform careful analysis
taking into account, in particular,  correlations of the  errors.

The triangles shrink for smaller values of $\sin^2  2\theta_{13}$ 
(Fig.~\ref{tri2}). 
According to Fig.~\ref{tri2}  which corresponds to $\sin^2  2\theta_{13}
= 0.03$,  for $\delta_D=90^{\circ}$  CP-violation might
be  established at $ \sim 2\sigma$ level. No conclusion can be made for 
$\delta_D < 70^{\circ}$. 

\begin{figure}[ht]
\centering\leavevmode
\epsfxsize=0.6\hsize
\epsfbox{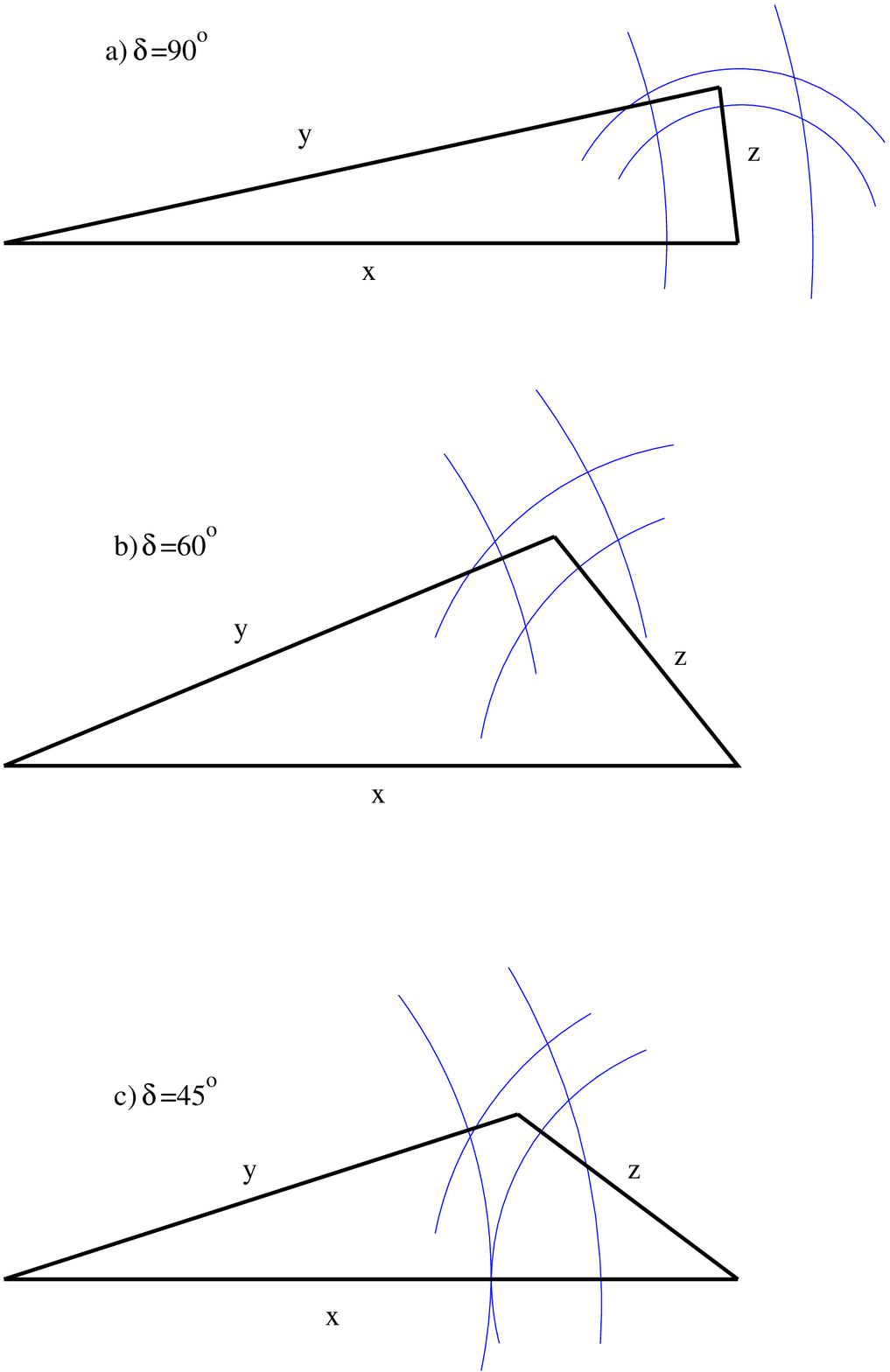}
\caption{
The same as  Fig. 1 for $\sin^2 2\theta_{13} = 0.18$ and $\tan 
\theta_{12}=1$.  
}
\label{tri3}
\end{figure}

The form of the triangle is also sensitive to variations of 
the angle $\theta_{12}$ within the allowed LMA region.  
In  Fig.~\ref{tri3}, we have set $\theta_{12}= {\pi \over 4}$ and 
$\sin^2 2\theta_{13}=0.18$. 
As follows from this figure with 10\% uncertainty in determination of the
sides, CP-violation can be established for $\delta_D=90^{o}$ and  
$\delta_D = 60^{\circ}$.

Note that $y \sim O (1)$ and $z$ is the smallest side, although 
its length may not be much smaller than others. 
So,  CP-violation implies  that 
\be
|U_{e1} U_{\mu 1}^*| <  |U_{e2} U_{\mu 2}^*| + |U_{e3} U_{\mu 3}^*|.
\label{ineq-tri}
\ee 
Similar triangles can be
obtained for the LOW and VAC solutions. 
The unitarity triangle is different in the case of the SMA-MSW solution. 
Taking $\tan^2 \theta_{12} = \tan^2 \theta_{sun} = 0.0016$, 
  $\sin \theta_{23} = 1/\sqrt{2}$ and  
$\sin \theta_{13} = 0.15$,  we find $y = 0.25$, $z = 0.96$. 
Now $y$ is the smallest side, are the two other sides have comparable  
lengths.  Note that in spite of small mixing of the electron neutrino  
the smallest side is not very small. Even in this case  
a moderate accuracy in determination of the sides would allow us to 
establish  CP-violation. 


In general, to establish  CP-violation, one  needs to construct the
triangle without using the  unitarity conditions. 
However, if we assume  that only three neutrino species take part
in the mixing and that there are no other sources of CP-violation apart
from the MNS-matrix, 
we can use some equalities which follow from  unitarity. 
In particular,  we can use the independent 
normalization conditions: 
\be
\sum_{i = 1,2,3} |U_{ei}|^2 = 1 ~,  ~~~~
\sum_{i = 1,2,3} |U_{\mu i}|^2 = 1. 
\label{normalization}
\ee 
In this case, the CP-violation effect can be mimicked 
at some level by  the 4th (sterile) neutrino. To eliminate such a  
possibility,  
one should check the normalization conditions experimentally.

Thus, to find the sides of the triangle  we should determine 
moduli of  four mixing matrix elements: 
\be
|U_{e2}|, ~~~~ |U_{\mu 2}|, ~~~~ |U_{e3}|, ~~~~ |U_{\mu 3}|.
\label{elements}
\ee 
They immediately determine the second and  third sides. The  
two other elements, $|U_{e1}|$ and  $|U_{\mu 1}|$,    
and consequently  the first side,  
can be found  from the normalization conditions (\ref{normalization}). 
For   the first side we have  
$|U_{e1}^*U_{\mu 1}| = 
\sqrt{(1 - |U_{e2}|^2 - |U_{e3}|^2)(1 -  |U_{\mu 2}|^2 - 
|U_{\mu 3}|^2)}$. 
Taking into account this correlation in determination of the 
sides of the triangle one  can estimate   
accuracy of  measurements of the elements (\ref{elements}) 
needed to establish 
CP-violation via the inequality (\ref{ineq-tri}). 
Let us introduce
\be
A \equiv |U_{e2}||U_{\mu2}| + |U_{e3}||U_{\mu3}|
-\sqrt{(1 - |U_{e2}|^2 - |U_{e3}|^2)(1 - |U_{\mu 2}|^2 - |U_{\mu 3}|^2)}
\ee
which is a measure of  CP violation. CP is conserved 
if  $A = 0$. For the most optimistic cases,  
where $U_{e3}$ is close to  the CHOOZ bound and  $\delta_D = 90^{\circ}$,  
we find  $A = 0.10 - 0.13$.

Suppose the elements $|U_{\alpha i}|$ are measured with accuracies
$\Delta |U_{\alpha i}|$.
Assuming that the errors $|\Delta U_{\alpha i}|$ are uncorrelated, 
we can write the error in the determination of $A$ as
\be
\Delta A = \sqrt{\sum_{\alpha = e,\mu, i = 2, 3} 
\left(\frac{dA}{d|U_{\alpha i}|}\right)^2
\left(\Delta |U_{\alpha i}|\right)^2}, 
\label{allcor}
\ee
where  
\be
\frac{dA}{d|U_{e2}|} = |U_{\mu 2}| +
\frac{|U_{e2}||U_{\mu1}|}{|U_{e1}|},~~~~~~~~~
\frac{dA}{d|U_{e3}|} = |U_{\mu 3}| +
\frac{|U_{e3}||U_{\mu1}|}{|U_{e1}|}, 
\label{e-el}
\ee
\be
\frac{dA}{d|U_{\mu2}|} = |U_{e 2}| + 
\frac{|U_{e1}||U_{\mu2}|}{|U_{\mu1}|},~~~~~~~~~
\frac{dA}{d|U_{\mu3}|} = |U_{e 3}| +
\frac{|U_{e1}||U_{\mu3}|}{|U_{\mu1}|}.
\label{m-el}
\ee 
As an  example,  let us choose the  oscillation parameters 
used in  Fig.~1 and $\delta_D = 90^{\circ}$.  
Then  from Eqs. (\ref{e-el}, \ref{m-el}) we find 
$dA/d|U_{e2}| = 0.82$,~  $dA/d|U_{e3}| = 0.77$,~  
$dA/d|U_{\mu2}| = 2.0$ and  $dA/d|U_{\mu3}| = 1.9$. 
Note  that  for muonic elements
the derivatives  are larger by  factor of  2. 
This is a consequence of the appearance of the 
relatively small element $|U_{\mu1}|$ in denominators 
of (\ref{m-el}).
So, the muonic  elements should be measured with the accuracy 
two times better than the electronic elements.

For our example we find  from  Eq. (\ref{allcor})  that $\Delta A < 0.065$, 
which would allow to  establish deviation of $A$ from zero at the 
$2\sigma$ level,  
if $\Delta |U_{e 2}| = \Delta |U_{e 3}| < 0.03$ and 
$\Delta |U_{\mu 2}| = \Delta |U_{\mu 3}| <  0.02$. 
This, in turn, requires the following upper bounds for relative accuracies 
of measurements of  the matrix elements:  $6\%$  for $|U_{e 2}|$, 
$17\%$  for $|U_{e 3}|$, and  $3\%$ for 
$|U_{\mu 2}|$ and  $|U_{\mu 3}|$.  
Since 
$$
\frac{\Delta |U_{\mu 2}|} {|U_{\mu 2}|}= \left(\frac{|U_{\mu 1}|}{|U_{\mu 2}|}\right)^2
\frac {\Delta |U_{\mu 1}|}{|U_{\mu 1}|} 
$$
and $U_{\mu 1} \simeq 0.5 U_{\mu 2}$, the required 
$3\%$   accuracy  in $|U_{\mu 2}|$  corresponds to
12 \% uncertainty in $U_{\mu 1}$. 
If there are  correlations between
$\Delta |U_{\alpha i}|$, the situation may  become  better.
So, the above estimations can be considered as  the  conservative ones.

\subsection{Present status}

At present, we cannot reconstruct the triangle:   knowledge of 
the mixing matrix is limited to the elements of the first row 
(from the solar neutrino data and CHOOZ/Palo Verde experiments) and the
third column (from the atmospheric neutrino data). To reconstruct 
the triangle one  needs to know at least one element from the block 
$U_{\beta i}$,  where $\beta = \mu, \tau,~~~  i = 1, 2$. That is, one
should measure the distribution of the $\nu_{\mu}$ (or/and
$\nu_{\tau}$) in the  
mass eigenstates with split by the solar $\Delta m^2$. Using the
unitarity
condition we can estimate only the ranges for these matrix elements. 
Clearly, present data are consistent with any value of the 
CP-violating phase and, in particular, with  zero value which 
corresponds to degenerate triangles.

Let us summarize our present knowledge of the 
relevant matrix elements. 

1). The values of  the mixing parameters $|U_{e1}|$ and $|U_{e2}|$ 
can be obtained from studies of solar neutrinos. 
Neglecting small effect due to $U_{e3}$, for the LMA-MSW 
solution we obtain
\be
\frac{|U_{e2}|}{|U_{e1}|} = |\tan \theta_{sun}| = 0.39 - 0.77, ~~~~~ 
(95 \% ~~ {\rm C.L.}) 
\label{solar} 
\ee
and then using the normalization condition:  
\be
|U_{e1}| \sim  [1  + \tan^2 \theta_{sun}]^{-1/2} =  0.79 - 0.93, ~~~~~ 
(95 \%  ~~ {\rm C.L.}) . 
\label{solar1}
\ee   

2). The absolute value of  $|U_{e3}|$  
is restricted from above by the  CHOOZ\cite{chooz} and Palo Verde
\cite{palo} experiments. 
The $2\nu$ analysis of the CHOOZ data \cite{chooz}, 
gives  for the best fit value of  $\Delta m_{atm}^2$   
\be
|U_{e3}| < 0.20,  ~~~~(90 \%  \ \ {\rm C.L.}).  
\ee
For lower values of $\Delta m_{atm}^2$, the bound is weaker: 
$|U_{e3}| < 0.22$.

3). The admixture of the muon neutrino 
in the third mass eigenstate, $|U_{\mu 3}|$, is determined by
the atmospheric neutrino data.   
Again, neglecting effects  due to non-zero $U_{e 3}$, 
we can write 
\be
4 |U_{\mu 3}|^2(1 -   |U_{\mu 3}|^2) = \sin^2 2\theta_{atm}, 
\ee     
where $\sin^2 2\theta_{atm}$ can be extracted, {\it e.g.}, from  
analysis of the zenith angle distribution of the  $\mu$-like events 
in terms of the $\nu_{\mu} -  \nu_{\tau}$ oscillations. Using the
Super-Kamiokande data, we find  
\be
|U_{\mu3}| = 0.707 ^{+ 0.12}_{-0.14}, ~~~~~ (90 \%   {\rm C.L.}). 
\label{bound}
\ee

4). At present,  there is no direct information
about  $|U_{\mu 1}|$ and  $|U_{\mu 2}|$. 
To measure these elements,   one needs to study the  oscillations 
of muon neutrinos driven by $\Delta m_{sun}^2$. 
The normalization condition  allows us to impose a bound on a combination
of these elements: 
\be 
 |U_{\mu 1}|^2 + |U_{\mu 2}|^2 = 1 - |U_{\mu 3}|^2 = (0.33 -
0.67).
\label{norm-con}
\ee
So, to determine $|U_{\mu 1}|$ and  $|U_{\mu 2}|$ 
separately we need to measure 
a combination of these elements which differs from the normalization 
condition (\ref{norm-con}).

\section{Reconstructing the unitarity triangle}

Let us consider the possibility to determine the triangle in the
forthcoming and future oscillation experiments. 
We suggest  a set of oscillation measurements  
with  certain configurations 
(base-lines, neutrino energies and features of detection) 
which will allow us to measure the moduli of the relevant matrix elements 
(see Eqs. (\ref{sides}, \ref{xyz})). 

In general, for $3 \nu$-system  the oscillation probabilities  
depend not only on moduli of the mixing matrix elements, 
(\ref{elements}) we are interested in,  but also
on other mixing parameters including the  unknown relative phases 
of the mixing matrix elements, $\delta_x$.  
Therefore, the problem is to select 
configurations of  oscillation measurements for which 
the dominant effect is determined by relevant moduli 
and corrections which depend on unknown elements and phases 
are negligible or sufficiently small.   

The hierarchy of mass splittings:  
$\Delta m_{atm}^2 \gg \Delta m_{sun}^2$ helps to  
solve the problem.  We  use 
\be
\epsilon \equiv \frac{\Delta m_{sun}^2}{\Delta m_{atm}^2} \sim 0.02 
\label{expansion}
\ee
as an expansion parameter,  
where the estimation corresponds to  the best fit values of 
the mass squired differences. Another small parameter in the problem is 
$|U_{e 3}|$.

In what follows,  we suggest a set of measurements
for which the oscillation probabilities depend mainly 
on the relevant moduli:  
\be 
P_{\alpha \beta} =  
P_{\alpha \beta}(|U_{e i}|,|U_{\mu i}|) + 
\Delta P_{\alpha \beta}  (\delta_x), ~~~~
\alpha, \beta = e, \mu~,  
\ee
where $\Delta P \ll P$.  We estimate corrections, 
$\Delta P_{\alpha \beta} (\delta_x)$, due to unknown mixing elements and
phases.

It is convenient to study the dynamics of oscillations 
in the basis of states obtained through 
rotation by the atmospheric mixing angle: 
$(\nu_e, \nu_{\mu}', \nu_{\tau}')$, where  
\be
\nu_{\mu}' = 
\frac{1}{\sqrt{1 - |U_{e 3}|^2}} (U_{\tau 3} \nu_{\mu} -
U_{\mu 3}\nu_{\tau}), ~~~~~~~~~~
\nu_{\tau}' =  \frac{1}{\sqrt{1 - |U_{e 3}|^2}} (U_{\tau 3}^*\nu_{\tau} 
+  U_{\mu 3}^* \nu_{\mu}).
\label{basis}
\ee
Projections of these states onto the mass eigenstates equal 
\be
\langle \nu_e | \nu_1 \rangle = U_{e1}^* , ~~~~
\langle \nu_{\mu}' | \nu_1 \rangle 
= - \frac{U_{e2}^*}{\sqrt{1 - |U_{e3}|^2}},  ~~~~
\langle \nu_{\tau}'| \nu_1 \rangle = 
- \frac{U_{e1} U_{e3}^*}{\sqrt{1 - |U_{e3}|^2}}  
\label{first}
\ee
and
\be
\langle \nu_e | \nu_3 \rangle = U_{e3}^*,~~~~
\langle \nu_{\mu}'| \nu_3 \rangle = 0, ~~~~
\langle \nu_{\tau}' | \nu_3 \rangle  = \sqrt{1 - |U_{e3}|^2}. 
\label{third}
\ee     
Note that in the limit $U_{e3} = 0$, the state  
$\nu_{\tau}'$ coincides with mass
eigenstate $\nu_3$, whereas 
$\nu_{\mu}' = -\sin \theta_{12}\nu_1 + \cos
\theta_{12}\nu_2$.

In matter, the system of three neutrinos 
$(\nu_e, \nu_{\mu}', \nu_{\tau}')$  has 
two resonances associated with the two different $\Delta m^2$. The
corresponding resonance energies for the typical density in the mantle of
the Earth,  are  
\be
E^{R}_{23} = 7 ~{\rm GeV}, ~~~~  E^{R}_{12} = 0.15 ~{\rm GeV}.  
\label{resener}
\ee
These energies determine the typical energy scales of the problem as well
as  the energies of  possible experiments.
Also  there are two length scales in the problem 
which correspond to the oscillation lengths: 
\be
l_{12} \equiv \frac{4 \pi E}{\Delta m_{sun}^2} = 5\cdot 10^{4}~ {\rm km} 
\left(\frac{E}{\rm GeV}\right), ~~~
l_{23}  \equiv \frac{4 \pi E}{\Delta m_{atm}^2} = 10^{3}~ {\rm km}
\left(\frac{E}{\rm GeV}\right)~. 
\label{length}
\ee
These  numbers  have been obtained  for   the best fit 
values of the mass squared differences.

Let us consider possibilities to determine the moduli of relevant 
elements  of mixing matrix (\ref{elements}) in turn. 

\subsection{$|U_{e3}^* U_{\mu 3}|$} 

In principle, this product can   be directly
measured in  studies of the $\nu_{\mu} - \nu_e$ oscillations driven by 
$\Delta m_{atm}^2$. Let us consider a relatively 
short baseline experiment in  vacuum.  
The transition probability can be written as 
\be
P_{\mu e} =  4|U_{e3}^* U_{\mu 3}|^2 \sin^2 
\frac{ \Delta m_{atm}^2 L}{4 E} + \Delta P_{\mu e},  
\label{irmi}
\ee
where $\Delta P_{\mu e}$ is the correction due to existence of 
the $\Delta m_{sun}^2$ splitting : $\Delta P_{\mu e} \rightarrow 0$ 
when $\Delta m_{sun}^2 \rightarrow 0$. 
Thus, if the original flux is composed of pure $\nu_\mu$ (or 
pure $\nu_e$), detecting  the  appearance of $\nu_e$ (or $\nu_\mu$), one 
can measure immediately $|U_{e3}^* U_{\mu 3}|$ provided that 
$\Delta P_{\mu e}$ is small enough.  
Note that $\Delta P_{\mu e}$ depends on mixing 
matrix elements $U_{\alpha 1}$, $U_{\alpha 2}$, ($\alpha = e, \mu$), 
both on their  absolute values and on  phases 
which are unknown. So, we cannot predict $\Delta P_{\mu e}$ 
and the only way to proceed is to find conditions for experiment at
which this value is small. 
An alternative method would be independent measurement of 
$|U_{e3}|$ and $|U_{\mu3}|$.

For neutrino energies, $E > 100$ MeV (which are of practical interest)
the oscillation length in vacuum, $l_{23}$, 
is more than several 
hundred kilometers. This means that the   experiment should be a
long-baseline one, and therefore oscillations will occur in the matter of
the Earth. 

In a medium with constant density\footnote{For simplicity we will
consider matter with constant density. Density  variation  
effects do not change our conclusions.} 
the probability can be written as 
\be
P_{\mu e} =   
\left|
(U_{e3}^m)^* U_{\mu 3}^m \left( e^{i\Phi^m_{32}} - 1 \right) + 
(U_{e1}^m)^* U_{\mu 1}^m \left(e^{i\Phi^m_{12}} -1 \right)   
\right|^2,
\label{prob-gen}
\ee
where   $U_{\alpha i}^m$ are  the mixing matrix elements 
in matter and $\Phi^m_{ij}$ is  the oscillation phase difference of  
$i-$ and  $j-$ eigenstates.  

In the {\it vacuum limit} (one may  consider a hypothetical configuration
of experiment where neutrino beam  propagates mainly in atmosphere or in a
tunnel), $U_{\alpha i}^m = U_{\alpha i}$ and $\Phi^m_i = \Phi_i$. 
The first term in (\ref{prob-gen}) corresponds to the  mode of
oscillation we are
interested in, and the second term is due to the $\Delta m_{sun}^2$ 
splitting. The main correction follows from the interference of these two
terms.

For the correction we find   
\be
\Delta P_{\mu e} \approx
- 2 \epsilon  \left|U_{e1}^* U_{\mu 1} U_{e3} U_{\mu 3}^*\right| \Phi_{32} 
[\sin(\delta_x 
- \Phi_{32}) - \sin \delta_x],    
\label{prob-cor}
\ee               
where $\delta_x$ is the unknown phase of the product of four mixing matrix
elements. In derivation of (\ref{prob-cor}), we have used the  
smallness of the phase $\Phi_{12}$: 
\be
\Phi_{12}  =   \epsilon \Phi_{32}, 
\label{phase-rel}
\ee
assuming that $\Phi_{32}  = O(1)$ (which maximizes the effect of
oscillations). Then the relative correction is of the order of
\be
\frac{\Delta P_{\mu e}}{P_{\mu e}} \sim
\epsilon \frac{\sin 2\theta_{sun}}{|U_{e3}|}.  
\label{corr-rel}
\ee
For the best fit values of the solar oscillation parameters (LMA-MSW 
solution)  
and $U_{e3} = 0.2$ we get $\Delta P_{\mu e}/P_{\mu e} \sim  0.1$. 
That is, the product $|U_{e3}^* U_{\mu 3}|^2$ can be measured  with 
accuracy not better than  10\% for maximal possible $U_{e3}$.  
Consequently, the accuracy in the determination of $|U_{e3}^* U_{\mu 3}|$ 
cannot be better than $5 \%$.  

There are two possibilities to  improve  the  accuracy: 
1) The main oscillation term and the interference term 
have different dependences  on $\Phi_{32}$ and therefore 
on $E/L$. So, in principle one can disentangle these terms by
studying the energy dependence of the effect. 2) The 
sign of the interference term can be changed  varying $E/L$. 
Therefore, the correction can be suppressed by averaging over energy, 
especially  if $\delta_x$ is small. 

Note that for other solutions of the solar neutrino problem 
(LOW, SMA, VAC),  $\Delta m_{sun}^2$ is much smaller and the correction 
is negligible.

In  the matter the dependence of the oscillation probabilities 
on mixing matrix elements becomes more complicated. 
However, there are two limits in which the dominant term of 
$P_{\mu e}$ can be reduced 
approximately to the form  (\ref{irmi}):  
(i) low energy limit $E \ll E_{13}^R$ in which  
matter corrections are small and (ii) short base-line 
limit $L \ll l_{13}^m$ where ``vacuum mimicking" condition  
is satisfied \cite{mimick}. 

Let us consider first the {\it low energy} case, $E \sim (200 - 500)$ MeV.  
The relative corrections due to matter effect to the 
main term in (\ref{irmi}) are of the order of
\be
\frac{l_{23}}{l_0} = 
\frac{2\sqrt{2} G_F n_e E}{\Delta m^2_{atm}}~,
\label{eps}
\ee     
where  $l_0 \equiv \sqrt{2} \pi/G_F n_e$ is the refraction length. 
For $E \sim 200$ MeV,  we have  $\epsilon \sim 0.02$,  while for 
$E \sim 1$ GeV,  the corrections reach  10\%. 
Moreover, the matter effect is of order 1  
for the correction term driven by  $\Delta m_{sun}^2$.

At low  energies the mixing in the heaviest eigenstate 
is only weakly affected  by matter, so that in the first approximation we
can take  $U_{e3}^m \approx  U_{e3}$,  $U_{\mu 3}^m \approx U_{\mu 3}$. 
For  $E \sim 200$ MeV  the oscillation length  due to $2 - 3$ level
splitting  is $\sim 200$ km, 
and therefore  the  optimal baseline would be $L \sim (100 - 200)$ km. 

The energies $E \sim 200$ MeV   are  in the  resonance interval 
for  $\Delta m^2_{sun}$. This means that the electron 
neutrino has comparable admixtures in the two 
light eigenstates:  $U_{e2}^m \sim U_{e1}^m \sim 1/\sqrt{2}$. 
The oscillation length  is of the order of the 
vacuum oscillation length (for the  LMA-MSW solution 
$l_{12} \sim 5 \cdot 10^{3}$ km), and therefore the oscillation phase
due to the 1-2 level splitting is small: 
$\Phi_{12}^m \sim 2\pi L/l_{12} \sim 
 \epsilon \ll 1$. 
These  features  simplify  the  analysis of the correction term.  
Indeed, using (\ref{basis}, \ref{first}, \ref{third})), we 
find    
\be
(U_{e1}^m)^* U_{\mu 1}^m \Phi_{12}^m \approx 
U_{e1}^*U_{\mu 1} 
\Phi_{12}. 
\label{u1e1}
\ee 
Therefore,  the relative correction  
$\Delta P_{\mu e}/P_{\mu e}$  which appears due to the interference 
of the term (\ref{u1e1}) with $(U_{e3}^*)^m U_{\mu 3}^m$  in
(\ref{prob-gen}) can be written as 
\be
\frac{\Delta P_{\mu e}}{P_{\mu e}} \sim
\epsilon 
\frac{\sin 2\theta_{sun}}{|U_{e3}|} 
\frac{|U_{\tau 3}|}{|U_{\mu 3}|} 
\approx 
\epsilon \frac{\sin 2\theta_{sun}}{|U_{e3}|}.  
\label{corr-rel-l}
\ee  
Here we have taken into account the relation between phases 
(\ref{phase-rel}). 
So, the expression for the correction is basically reduced 
to  that in the  vacuum oscillation case given 
by Eq. (\ref{corr-rel}). 

Similar considerations hold for the {\it antineutrino} channel.
We find that not only the main term but also the orders of 
magnitude of the corrections 
coincide with those for the vacuum case.

Let us consider the case of {\it  high} energies,  
$E \sim  (5 - 10)$ GeV,   
and relatively short baselines, $L \sim 700$ km, 
for which the ``vacuum mimicking" condition, $L \ll l_{13}^m$, 
is satisfied. Notice that  these  energies are  
in the range of the resonance due to $\Delta m_{atm}^2$ (\ref{resener}).

In the limit where the   $\Delta m_{sun}^2$ splitting can be neglected the 
probability (for the constant density case) becomes   
\be
P_{\mu e} = \frac{4|U_{e3}^* U_{\mu 3}|^2}{R^2_{13}} 
\sin^2 \frac{1}{2} \Phi_{32}  R_{13}  + \Delta P_{\mu e},
\label{prob-em}
\ee     
where  
\be
R^2_{13} \equiv \left(\cos 2\theta_{13} - \frac{l_{23}}{l_{0}}\right)^2 
+ \sin^2 2\theta_{13}. 
\label{resfac}
\ee 
The oscillation phase in matter, $\Phi_{32}^m$, can be written  in terms
of
the phase in vacuum, $\Phi_{32}$,  as 
\be
\Phi_{32}^m = \Phi_{32} R_{13}, ~~~~~ \Phi_{32} \equiv  2\pi L/l_{32}.    
\label{phase}
\ee
For $\Phi_{32}^m \ll 1$ the expansion of the probability
(\ref{prob-em}) in powers of $\Phi_{32}^m$ leads to 
\be
P_{\mu e} =  4|U_{e3}^* U_{\mu 3}|^2 \Phi_{32}^2  
\left(1 - \frac{R_{13}^2}{3} \Phi_{32}^2 \right). 
\label{prob-exp}
\ee    
Note that the first (leading)  term in (\ref{prob-exp}) 
reproduces the vacuum probability (vacuum mimicking).  

The correction depends on energy. 
In the resonance $R^2_{13} = \sin^2 2\theta_{13}$, 
and  therefore  according to (\ref{prob-exp}) 
corrections are below 10\%,  even if $\Phi_{32}^2 \sim 1$. 
Below the resonance:  for $l_{23}/l_0 \rightarrow 0$, 
we have $R^2_{13} \rightarrow 1$  and 
to have a small correction  $\Phi_{32}^2$ should be small, thus leading to 
suppression of the 
oscillation effect. Therefore, for such a type of experiment, the optimal 
range of energies  is $E = (5 - 10)$ GeV, and the optimal baseline $\sim 
700$
km. Above the resonance where $R^2_{13} \rightarrow \infty$, the 
oscillation effect is  even more suppressed.

Let us evaluate the correction to (\ref{prob-exp}),    
$\Delta P_{\mu e}$, due to the solar mass splitting. 
For high energies both oscillation phases $\Phi_{12}^m$ and 
$\Phi_{32}^m$ are small, so that from Eq. (\ref{prob-gen}) we obtain
the interference term 
\be
\Delta P_{\mu e} =
2 Re \left[ (U_{e3}^m)^* U_{\mu 3}^m \Phi^m_{32} 
U_{e1}^m (U_{\mu 1}^m)^* \Phi^m_{12} \right].
\label{prob-cor5}
\ee        
Considering explicitly the mixing in the $\nu_e - \nu_{\tau}'$ 
system we find    
\be
(U_{e3}^m)^* U_{\mu 3}^m \Phi^m_{32} \approx U_{e3}^* U_{\mu 3} \Phi_{32}. 
\label{rel1}
\ee 
The mixing of the electron neutrino in the first 
eigenstate is strongly suppressed by matter effect: 
\be
U_{e1}^m (U_{\mu 1}^m)^* \approx U_{e1} U_{\mu 1}^*
\frac{l_0}{l_{12}}~ ,   
\label{rel2}
\ee    
where ${l_0}/{l_{12}} \ll 1$.  
As follows from the level crossing scheme, in the resonance region  
the phase difference between the two light eigenstates is 
\be
\Phi_{12}^m \approx - \Phi_{32}.   
\label{menhaa}
\ee
Plugging  the expressions (\ref{rel1}), (\ref{rel2}) and 
(\ref{menhaa}) into Eq. (\ref{prob-cor5}) we find 
\be
\Delta P_{\mu e} \approx
2 Re \left[U_{e3}^* U_{\mu 3} U_{e1} U_{\mu 1}^*\right]
\Phi_{32}^2 \frac{l_0}{l_{12}}.  
\label{prob-cor6}
\ee  
The  relative correction can then be written  as 
\be
\frac{\Delta P_{\mu e}}{P_{\mu e}} \approx  \epsilon\frac{\sin
2\theta_{sun}}{|U_{e3}|}
\frac{E^{R}_{23}}{E}.   
\label{corr-rel7}
\ee               
Near the resonance the relative correction is similar to that in the low 
energy limit or in vacuum case. 
The correction can be further suppressed if $E > E^R_{13}$. 
At the same time,  
with the increase of energy, the phase of oscillations ($\Phi_{32}$ in Eq. 
(\ref{prob-exp})) decreases,  
and therefore  the number of events decreases quadratically.

Let us consider the antineutrino channel. At high energies, 
$\bar{\nu}_{2m} \approx \bar{\nu}_{\mu}'$ 
and $\bar{\nu}_e$ mixes with $\bar{\nu}_{\tau}'$ in the $\bar{\nu}_{1m}$
and $\bar{\nu}_{3m}$ eigenstates. 
In the limit  $\bar{\nu}_{2m} =  \bar{\nu}_{\mu}'$,  the 
 standard vacuum expression for the probability 
$P_{e\mu}$ (Eq. (\ref{prob-exp}) with $R_{13} 
\rightarrow \bar R_{13}$) is reproduced. 

The correction is related to the admixture  of $\bar{\nu}_e$ 
in $\bar{\nu}_{2m}$ which is determined by $\sin \theta_{12}^m$ and 
is strongly suppressed by matter.
A straightforward calculation give 
\be 
(\bar{U}_{e1}^m)^* \bar{U}_{\mu 1}^m \approx 
- \frac{\bar{U}_{\mu3} \bar{U}_{e3}^*}{\bar R_{13}}
+ \bar U_{e1}^*
\frac{U_{\mu 1}}{\bar R_{12}}
\frac{\cos\theta_{13}^m}{\cos \theta_{13}},
\label{bahs}
\ee  
where the correction is given by the second term. 
Taking into account that $\Phi_{12}^m = \Phi_{12} R_{12} 
= - \Phi_{32} R_{12} \epsilon$ 
one can find  
\be
\frac{\Delta \bar{P}_{\mu e}}{\bar{P}_{\mu e}} \approx 
\epsilon 
\frac{\sin 2\theta_{sun}}{U_{e3}}
\frac{|\bar{U}_{\tau 3}|}{|\bar{U}_{\mu 3}|} 
\frac{\cos\theta_{13}^m}{\cos \theta_{13}}~
\approx
\epsilon \frac{\sin 2\theta_{sun}}{|U_{e3}|}.  
\label{corr-rel8}
\ee 
Note that in contrast to the neutrino case the correction does  
not change with energy. So, using neutrino beam  seems to be more  
promising  because for neutrinos  relative corrections  decrease
with increase of energy.  
 
We now discuss the sensitivity of 
upcoming and  planned  experiments to the product 
$|U_{e3}^* U_{\mu 3}|$.  It  was shown \cite{barger} that  
combining the data from the MINOS and  ICARUS experiments, one can   
obtain an upper bound  $\sin^2 2 \theta_{13} < 0.01 $  at the 
95\% C. L.  This would correspond to 
$|U_{e3}^* U_{\mu 3}| < (0.03 - 0.04)$ which is about 
4 times stronger than the present bound: $|U_{e3}^* U_{\mu 3}| < 0.15$.  
The  searches for the $\nu_{\mu} - \nu_{e}$ oscillation will 
be performed in  
phase I of the JHF project. The sensitivity to the product 
$|U_{e3}^* U_{\mu 3}|$ can reach to 0.02 \cite{jhf-kamioka}. 
Therefore, if $|U_{e3}^* U_{\mu 3}|$ is at the border of 
the present upper bound it will be measured with about 15 \% accuracy. 
Neutrino factories will be sensitive to  
$|U_{e3}^*U_{\mu 3}|$ down to $10^{-3}$. 
However,  for   $|U_{e3}^*U_{\mu 3}| <  {\rm few} \times 10^{-3}$  
the correction due to non-zero value of $\Delta m_{sun}^2$ 
(in the case of  LMA-MSW solution)   
will be comparable with the main term (see \cite{irina} for related 
discussion). 
For other solutions the corrections are negligible.

\subsection{$|U_{e3}|$}  

Independent determination of $|U_{e3}|$ seems to be important in 
view  of the  difficulties associated with 
the direct measurements of $|U_{e3}^* U_{\mu3}|$ discussed 
in sect. 3.1. 
Knowledge of $|U_{e3}|$ is also needed for a precise determination of  
$|U_{e1}|$,  $|U_{e2}|$ and other mixing elements. 

The survival probability for $\nu_e$-oscillations in vacuum can be written
as 
\be
P_{e e} =
\left| |U_{e3}|^2 \left( e^{i\Phi_{32}} - 1 \right) +1+
|U_{e1}|^2 \left(e^{i\Phi_{12}} -1 \right)
\right|^2. 
\label{prob-ee}
\ee    
Note that,  in contrast to the conversion case, the  
probability amplitude depends on the required moduli of the
matrix elements.  A similar  analysis holds for antineutrinos.  

For low (reactor) energy experiments the matter effects are negligible 
and the probability equals 
\be
P_{e e} = 1 - 4 (1 - |U_{e3}|^2) |U_{e3}|^2 \sin^2
\frac{\Phi_{32}}{2} + \Delta P_{e e}~.
\label{probab-ee}
\ee  
Here the correction $\Delta P_{e e}$ due to the $\Delta m_{sun}^2$ 
splitting can be evaluated as  
\be
\Delta P_{e e} = 2|U_{e1}|^2 |U_{e3}|^2 
\Phi_{12} \sin\Phi_{32} - \frac{1}{4} \Phi_{12}^2 \sin^2 2\theta_{sun}.
\label{prob-eecor}
\ee  
The relative correction  is  small:  
$\Delta P_{e e}/P_{e e} < 2$ \%, so that in principle,   $|U_{e 3}|$  
can be determined  with better than 1\% accuracy. 
Experimental errors in the measurement of $P_{ee}$ will dominate. 

Let us comment on the experimental prospects for measuring 
$|U_{e3}|$. A new reactor experiment, Kr2Det,  has been proposed 
which will be able to set the bound $|U_{e3}|< 0.07$ at the 
90 \% C. L. \cite{kozlov}. 
This  bound can be used to  estimate the  sensitivity. 
If, {\it e.g.},   $|U_{e3}|=0.2$, one would expect that 
the experiment will give 
$|U_{e3}|^2 = 0.040 \pm 0.005$. Consequently, $|U_{e3}|$ itself 
will be determined with about 6\% accuracy. 
In fact,  the situation can be slightly better. If $|U_{e3}|$ 
is  near its upper bound,  one can study 
spectrum distortion and therefore to perform a more 
accurate determination  of  $|U_{e3}|$.

It is not clear if future measurements allow us to 
measure $|U_{e3}|$ precisely enough to reconstruct 
the third side of the triangle. But certainly, they will contribute to 
a more precise determination of $|U_{e1}|$ and $|U_{e2}|$.

\subsection{$|U_{\mu 3}|$} 

The present analysis of the atmospheric neutrino
data in terms of $2\nu$- mixing gives the bound (\ref{bound}) 
on $|U_{\mu  3}|$.  Note that at this accuracy, 
the effects of non-zero $U_{e 3}$ and subleading modes driven 
by $\Delta m_{sun}^2$ are unimportant.  
Indeed, in \cite{lisi}, it is shown that the allowed  region for 
$|U_{\mu 3}|$ does not change considerably as $|U_{e 3}|$ varies between
zero and its maximal possible value. 
To reconstruct the unitarity triangle, we need a more precise measurement 
for $|U_{\mu 3}|$, which requires $3\nu$ analysis taking into account the 
effect of non-zero $|U_{e3}|$.

The element $|U_{\mu3}|$  can  be  
measured in   $\nu_{\mu}$-disappearance
due to oscillations driven by $\Delta m_{atm}^2$. 
The $\nu_{\mu}$-survival probability in a uniform medium equals:  
\be
P_{\mu \mu}  =
\left| |U_{\mu 3}^m|^2 \left( e^{i\Phi_{32}^m} - 1 \right) + 1 +  
|U_{\mu 1}^m|^2 \left(e^{i\Phi_{12}^m} -1 \right)\right|^2.
\label{prob-mumug}
\ee    
Again, there are two limits in which the dominant term of 
this  probability reduces to the vacuum oscillation probability 
plus small corrections: 
(i) the low energy limit ($E \ll E_{13}^R$),  
(ii) and the high energy case ( $E \geq  E_R$) with a small baseline   for
which the vacuum mimicking condition is satisfied.

Let us consider first the {\it high energy} limit. The dynamics is
particularly simple, if $E >  E_{13}^R$. In this case 
$\nu_{1m} \approx \nu_{\mu}'$, whereas the  states $\nu_e$ and 
$\nu_{\tau}'$ strongly mix in $\nu_{2m}$ and $\nu_{3m}$.
For the  phases  we have  
$\Phi_{12}^m \approx \Phi_{32}$ and 
$\Phi_{32}^m \approx \Phi_{32} R_{13} \ll 1$.  
Neglecting the admixture of $\nu_{\tau}'$ in $\nu_{1m}$ and $\nu_{2m}$ 
(which is smaller than $|U_{e3}| \epsilon$)  we obtain
\be
|U_{\mu 1}^m|^2  \approx   \frac{|U_{\tau 3}|^2}{1 - |U_{e3}|^2}, ~~~~ 
|U_{\mu 3}^m|^2  \approx  \frac{|U_{\mu 3}|^2}{1 - |U_{e3}|^2} 
\cos^2 \theta_{13}^m~, 
\label{sss}
\ee
where $\cos^2 \theta_{13}^m = 
[1 + (\cos 2 \theta_{13} - l_{13}/l_0)/R_{13}]/2$. 
Inserting these matrix elements into 
(\ref{prob-mumug}) we can reduce the probability to the form 
\be
P_{\mu \mu} = 1 - (1 - |U_{\mu 3}|^2) |U_{\mu 3}|^2 
\Phi_{32}^2  + \Delta P_{\mu \mu}~,
\label{prob-mm}
\ee 
where $\Delta P_{\mu \mu}$ is the correction due to 
matter effects, non-zero value of $U_{e 3}$ and $\Delta m_{sun}^2$.
Recalling $\Phi_{32}^m \ll 1$, we can write 
\be
\Delta P_{\mu \mu} = 
\left[|U_{e3}|^2 |U_{\mu 3}|^2  - 
2|U_{\mu 3}|^2 (1 - |U_{\mu 3}|^2)(R_{13} \cos^2 \theta^m_{13} - 
|U_{e3}|^2)  
\right] \Phi_{32}^2 . 
\label{corr-mumu}
\ee   
At the resonance, $R_{13} \cos^2 \theta^m_{13} \approx |U_{e3}|$ 
and the corrections are strongly suppressed. 
However the resonance region is rather narrow.  
Above  the resonance, $R_{13}  \cos^2 \theta^m_{13} \approx 
|U_{e3}|^2/R_{13}$, and  
as follows from (\ref{corr-mumu}),   the relative corrections 
 can be estimated as 
$\Delta P_{\mu \mu}/ P_{\mu \mu} \sim |U_{e3}|^2 < 0.04$. Moreover, 
the corrections are calculable and can be  taken into account 
once $|U_{e3}|^2$ is measured  even with  a reasonable  accuracy. 
Corrections, which depend on $U_{\mu 1}$ and $U_{\mu 2}$,  
are suppressed by the ratio $\epsilon$ at $E \sim E^R_{13}$.

In the antineutrino channel we find 
\be
|\bar{U}_{\mu 3}^m|^2  \approx  \frac{|U_{\mu 3}|^2}{1 - |U_{e3}|^2}
\left(1 - \frac{|U_{e 3}|^2}{a_{13}^2} \right),
\label{u3mbar}
\ee     
where $a_{13}\approx 1 - 2 |U_{e 3}|^2 + l_{13}/l_0$, 
$\bar{\Phi}_{32}^m \approx \Phi_{32}$, and 
\be
|U_{\mu 1}^m|^2  \approx   
\frac{|U_{\mu 3}|^2}{1 - |U_{e3}|^2} 
\left[\frac{|U_{e 3}|^2}{a_{13}^2}
- \frac{|U_{\tau 3}|}{|U_{\mu 3}|}\frac{\sin 2\theta_{sun}}{\bar{R}_{12}} 
\frac{|U_{e 3}|}{a_{13}} \cos\delta_x \right].  
\label{u1mbar}
\ee     
Here $\delta_x$ is the unknown phase of the product of matrix 
elements. 
We can estimate $\bar{\Phi}_{12}^m = 
\bar{\Phi}_{13}^m - \bar{\Phi}_{23}^m \approx
\bar{\Phi}_{13}^m + \Phi_{32} \approx   
{\Phi}_{32}(1 -  \bar{R}_{13})$. 
Using the first term in the Eq. 
(\ref{u1mbar}) and  keeping  the lowest order terms in   
$|U_{e 3}|^2$, we find 
\be
\bar{P}_{\mu \mu} = 1 - \Phi_{32}^2
\left( 
(1 - |U_{\mu 3}|^2) |U_{\mu 3}|^2
+ |U_{e3}|^2 |U_{\mu 3}|^2 
(1 - a_{13}^{-2})(1 + a_{13}^{-2} -  2|U_{\mu 3}|^2)
\right). 
\label{prob-mmbar} 
\ee 
As in the neutrino case, we  assumed  $\bar{\Phi}_{13}^m \ll 1$.
According to (\ref{prob-mmbar}) 
the correction to the standard 
expression for $2\nu$ probability (the first term in the brackets) is of the  order 
of  $|U_{e3}|^2$ with coefficient smaller 
than 1. Note that, the unknown phases are not involved in the  Eq. 
(\ref{prob-mmbar}), so the corrections will be  calculable  
once $|U_{e3}|$ is measured.  

The relative corrections which depend on unknown phases originate from the
interference of the term proportional to $|\bar U_{\mu 1}^m|^2$ (the third 
term in Eq. (\ref{prob-mumug}) replacing $\Phi_{ij}^m\rightarrow \bar 
\Phi_{ij}^m$ and $|U_{\mu i}^m|\rightarrow |\bar U_{\mu i}^m|$)
and the main term. They can be estimated as 
\be
\frac{\Delta \bar{P}_{\mu \mu}}{1-\bar{P}_{\mu \mu}} 
\sim
2 \epsilon \sin 2\theta_{sun} |U_{e3}|.
\label{corr-rel9}
\ee
Note that in contrast to the case of determination of 
$|U_{e3}^* U_{\mu 3}|$, 
the relative corrections are suppressed by $|U_{e3}|$ because, in this case, 
the main term is larger; it corresponds to  
the  dominant mode of oscillations ({\it i.e.,} $P_{\mu \mu}\sim 1$, while 
$P_{\mu e}\ll 1$). 

Let us consider the {\it low energy}  experiment with $E \sim E^R_{12}
\sim (200 - 500)$ 
MeV. In this case the $\Delta m_{atm}^2$-driven oscillations are in the 
quasi-vacuum regime ($U_{\mu 3}^m \approx U_{\mu 3}$, $\Phi_{32}^m \approx 
\Phi_{32}$) and the base-line can be relatively small: 
$L \sim l_{12} \sim 100$ km. On the other hand, the oscillations driven 
by $\Delta m_{sun}^2$ are in the vacuum mimicking regime: 
$\Phi_{12}^m \ll 1$. It can be  shown that,
\be
|U_{\mu 1}^m|^2  \approx \frac{|U_{\tau 3}|^2}{1 - |U_{e3}|^2}  
\sin^2 \theta_{12}^m, ~~~~
\Phi_{12}^m \approx R_{12} \Phi_{12} = \epsilon R_{12} \Phi_{13}, 
\label{ssslow}
\ee
where $\sin^2 \theta_{12}^m = [1 - (\cos 2 \theta_{12} -
l_{12}/l_0)/R_{12}]/2$, and $R_{12}$ is the resonance factor 
for the (1 - 2) system:  
\be
{R}_{12} = \sqrt{\left(\cos 2 \theta_{12} - \frac{l_{12}}{l_0} 
\right)^2 + \sin^2 2\theta_{12} }.  
\label{r12}
\ee
Inserting the matrix element (\ref{ssslow})  into
(\ref{prob-mumug}), we can  reduce the probability to 
the form of Eq. (\ref{prob-mm}) with 
\be
\Delta P_{\mu \mu} =
2 \epsilon |U_{\tau 3}|^2 |U_{\mu 3}|^2  
\Phi_{32} \sin \Phi_{32}   
R_{12} \sin^2 \theta_{12}^m~. 
\label{corrolation-mumu}
\ee
Let us consider last two factors  in this expression. 
In the resonance, $R_{12} \sin^2 \theta^m_{12} \approx 
R_{12}/2 = \sin 2\theta_{sun}/2$,  
but above  it  
$R_{12} \sin^2 \theta^m_{12} \rightarrow l_{12}/ l_0$ 
and the correction increases with energy too. 
Below the resonance $R_{12} \sin^2 \theta^m_{12} 
\rightarrow 
\sin^2 \theta_{sun}$. 

Thus,  in the resonance region and below it, the correction is small and 
of the order of $\Delta m_{sun}^2 /\Delta m_{atm}^2$. 
The corrections due to admixture of $\nu_{\tau}'$ in the lowest  mass
eigenstate (which we have neglected) are of the order 
$|U_{e3}|^2$.  As in the high energy limit, the relative corrections are
restricted to 
$\Delta P_{\mu \mu}/ P_{\mu \mu}  < 0.04$, and moreover, the  dominant 
part of
these corrections can be calculated in terms of $|U_{e3}|$. 

In the antineutrino channel, similar consideration 
gives the following corrections 
which can be calculated in terms of the moduli of the matrix elements: 
\be
\Delta \bar{P}_{\mu \mu} = 
\epsilon 
\frac{|U_{\tau 3}|^2 |U_{\mu 3}|^2}{(1 - |U_{e3}|^2)^2}
\Phi_{32} \sin \Phi_{32}  
(\bar R_{12} - \sqrt{\bar R_{12}^2 - \sin^2 
2\theta_{sun}}).
\label{corr-mumubar}
\ee  
The relative corrections are of the order of $\Delta m_{sun}^2/\Delta
m_{atm}^2$. 
The corrections which depend on unknown phases, are further suppressed
($ \sim |U_{e3}| \epsilon$).

Studying the disappearance of $\nu_\mu$, the MINOS
experiment will determine $\Delta m_{atm}^2$ and $(1 - |U_{\mu 3}|^2)
|U_{\mu 3}|^2$ with 10 $\%$ accuracy 
at the  $99\%$ C.L. after  10 kton-years of 
data taking \cite{minos,barger}. Much higher precision can be achieved in
phase I of JHF: the oscillation parameters  
($(1 - |U_{\mu 3}|^2)|U_{\mu 3}|^2$ and $\Delta m_{atm}^2$) will be 
determined  with  $1\%$ uncertainty \cite{jhf-kamioka}. 
Thus,  there are good perspectives to determine $|U_{\mu3}|$ with 
precision better than 2 - 4 \%. 

Notice that the future atmospheric neutrino experiment, MONOLITH,
can measure  $\sin^2 2\theta_{23}$   with
uncertainty of 8\% \cite{monolith}.

\subsection{$|U_{e1}|$ and $|U_{e 2}|$}

The values of $|U_{e1}|$ and $|U_{e 2}|$ can be obtained from  
the solar neutrino data. 
To first approximation, due to the low energies of solar neutrinos  
the  matter effect  on  $|U_{e3}|$ is negligible 
and the solar neutrino conversion  driven by $\Delta m_{atm}^2$  will 
produce only an averaged oscillation effect.  
In this case the survival probability 
equals \cite{lim} 
\be
P_{e e} =  (1 - |U_{e3}|^2)^2 P_2 (\tan^2\theta_{sun}, \Delta m_{sun}^2) 
+ |U_{e3}|^4,
\label{prob-sun}
\ee 
where 
\be
\tan^2\theta_{sun} = \frac{|U_{e2}|^2}{|U_{e1}|^2}  
\label{tan-sun}
\ee
and $P_2$ is the two neutrino oscillation (survival) 
probability determined   
from the solution of the two neutrino 
$(\nu_e' - {\nu}_e)$ evolution equation with the  
oscillation parameters $\tan^2\theta_{sun}$, $\Delta m_{sun}^2$ 
and the effective potential $(1 - |U_{e3}|^2)V_e$.

Precise measurements of $|U_{e3}|^2$ will be performed by   
the KamLAND experiment 
for which $P_{ee}$ is given by (\ref{prob-sun}) with  
$P_2$ determined by the oscillation formula in vacuum: 
\be
P_2 = 1 - \frac{4|U_{e1}|^2|U_{e2}|^2}{(1 - |U_{e3}|^2)^2} \sin^2
\frac{\Phi_{12}}{2}~. 
\label{p2}
\ee
The expected error in determination of  $\sin^2 2\theta_{sun}$, 
and therefore the combination $|U_{e1}|^2|U_{e2}|^2$
is around 5\% \cite{kamland}. Then using  the measured value of 
$|U_{e1}||U_{e2}|$  and the normalization condition, 
$|U_{e1}|^2 + |U_{e2}|^2 = 1 - |U_{e3}|^2$,~ we can find 
$|U_{e1}|$ and $|U_{e2}|$, separately. The accuracy can be better than 
(2 - 3)\% .

\subsection{$|U_{\mu1}|$ and $|U_{\mu2}|$} 

The determination of $|U_{\mu1}|$ and $|U_{\mu2}|$ is the most challenging
part of the method. Note that in contrast to
$|U_{e3}^* U_{e\mu}|$ (see sect. 3.1),  it is not possible to measure 
the combinations $|U_{e1}^* U_{\mu 1}|$ or $|U_{e2}^* U_{\mu2}|$, 
directly from the oscillation experiments. 
Indeed, in vacuum the $\nu_{\mu} - \nu_e$ transition probability 
is determined by the product  
$Re\left[U_{\mu1}^* U_{e1} U_{\mu2} U_{e2}^*\right]$ 
which depends not only on the absolute values 
of the matrix elements but also on their phases. 
(For example, in the case that $\Delta  m_{sun}^2 L/ E$ 
is not resolved, the probability $P_{e \mu}$ is determined by the 
combination $|U_{\mu1}^*U_{e1}+U_{\mu 2}^*U_{e2}|$.)
Therefore we will consider the possibility to measure 
separately $|U_{\mu1}|$ and $|U_{\mu2}|$, so that
the second side of the triangle can be found using the 
electron matrix element $|U_{e2}|$ obtained in  
other  experiments. In fact,  it is sufficient to measure some
combination of $|U_{\mu1}|$ and $|U_{\mu2}|$ which differs 
from the normalization condition (\ref{norm-con}). 
This requires an experiment 
sensitive  to  the splitting between the first and 
second levels associated with $\Delta m_{sun}^2$   which appears 
usually as a subdominant mode. To suppress the  leading effect 
and the interference of the leading and sub-leading modes, the
oscillations 
driven by $\Delta m_{atm}^2$ should be averaged out. 
This condition necessitates  the following experimental configuration: 

1). The energy of  beam  should be low: $E < 1$ GeV.  

2). The baseline should be large:  $L\gg  l_{23}$
(in contrast to  configurations considered in the previous subsections). 
Moreover, to avoid suppression of the subdominant mode we need 
$L$ to be of the order of the  oscillation length due to the (1 - 2) 
splitting. 

At $E < 0.5$ GeV, we have $l_{23}  \sim 500$ km, and  consequently, 
to reach averaging  the
baseline can
be  $\sim 2000$ km. In this case $\Phi_{12}^m \sim O(1)$.

To produce muons, we need   
$E > 100$ GeV. For these energies matter effects on (1 - 2) mixing 
are non-negligible and 
moreover, since the baseline is large, no 
vacuum mimicking will occur.

The   experiment we have arrived 
at,  seems even more difficult  than  that   
for direct measurements of the CP-asymmetries 
\cite{jhf-kamioka}.   
However, our proposed experiment  measures quantities  
different  from asymmetries, and moreover, only one beam, neutrino or
antineutrino, is sufficient.

Let us consider the $\nu_{\mu} - \nu_{\mu}$ oscillation 
({\it disappearance}) experiment with 
$E \sim E^R_{12} \sim (200 - 500)$ MeV and $L \geq 2000$ km.   
At these energies the influence of the matter effect on flavor 
mixing in the third mass eigenstate 
is small so that we can  take $U_{e3}^m \approx U_{e3}$  
and also $U_{\mu 3}^m \approx U_{\mu 3}$. 
(The corrections are of  order of $\epsilon$.)
Therefore, the normalization condition gives $|U_{e1}^m|^2 + 
|U_{e2}^m|^2 = 1 -  |U_{e3}^m|^2  \approx 1 -  |U_{e3}|^2$. 
The mixing is reduced to the mixing in $2\nu$-system, 
so that matrix elements in matter can  be obtained by substituting 
$U_{e1} \rightarrow U_{e1}^m$,   $U_{e2}  \rightarrow U_{e2}^m$.

The  general form of the probability in a  
medium with constant density is given by Eq. (\ref{prob-mm}). Let us 
calculate $|U_{\mu 1}^m|$. 
First  we express the vacuum  value of $U_{\mu 1}$ in terms of 
$U_{e1}$, $U_{e2}$ and mixing in the third state, $U_{\alpha 3}$ 
$(\alpha = e, \mu)$. To do this, we use 
$U_{\mu 1} = \Sigma_{\alpha} \langle \nu_{\mu}|\nu_{\alpha}'\rangle
\langle \nu_{\alpha}'| \nu_1 \rangle$ and the relations
(\ref{basis}, \ref{first}, \ref{third}). 
Then  in the expression for  
$U_{\mu 1} =  U_{\mu 1}(|U_{e1}|, |U_{e2}|, |U_{\alpha 3}|)$ we
substitute $U_{ei} \rightarrow U_{ei}^m$ $(i = 1, 2)$. 
A straightforward    calculation gives 
\be
U_{\mu 1}^m = -\frac{1}{1 - |U_{e3}|^2} 
\left[|(U_{e2}^m)^* U_{\tau3}|   +  U_{e1}^m U_{e3}^* U_{\mu 3} 
\right]. 
\label{umu1}
\ee
Mixing elements in matter can be written as
\be
|U_{e1}^m|^2 = |U_{e1}|^2 \frac{\cos^2\theta_{12}^m}{\cos^2\theta_{12}} = 
\frac{|U_{e1}|^2}{R_{12}} \frac{R_{12} + \cos 2 \theta_{12} -
l_{12}/l_0}{1 + \cos 2 \theta_{12}}
\label{ue1}
\ee
(here $\cos 2 \theta_{12} \equiv 2 |U_{e1}|^2/(1-|U_{e3}|^2)  - 1$),  
\be
|U_{e2}^m|^2 = 1 - |U_{e1}^m|^2  - |U_{e3}|^2 
\label{ue2}
\ee
and
\be
(U_{e1}^m)^* U_{e2}^m \approx \frac{1}{R_{12}} U_{e1}^* U_{e2}. 
\label{uemix}
\ee
Using these equations we can express $|U_{\mu 1}^m|$ 
in terms of  the mixing parameters in vacuum as 
\be
|U_{\mu 1}^m|^2 = \frac{1}{R_{12}} |U_{\mu 1}|^2 + F,  
\label{umu1sq}
\ee
where 
\be
F  =  \frac{1}{(1 - |U_{e3}|^2)}
\left[ |U_{\tau3}|^2 f_+  +  |U_{e3}|^2 |U_{\mu 3}|^2
f_-
\right]
\label{FFF}
\ee
and 
\be
f_\pm = \frac{R_{12} -1  \pm l_{12}/l_0}{2 R_{12}}. 
\label{fff}
\ee
Note that in the vacuum limit $f_\pm \rightarrow 0$, $R_{12} \rightarrow
1$
and $F \rightarrow 0$. At the resonance,  
$R_{12} \rightarrow \sin  2\theta_{12}$ and  
above the resonance where $E\gg E_R^{12}$
\be
|U_{\mu 1}^m|^2 \rightarrow   F \approx 
\frac{|U_{\tau 3}|^2}{1 - |U_{e3}|^2}.  
\label{mire}
\ee    
In  this case the dependence  of $|U_{\mu 1}^m|^2$,  and consequently 
of the probability, on $|U_{\mu 1}|$ disappears in agreement with our 
result for the high energy version of the experiment in sect. 3.3. 
The survival probability can be written as  
\be
P_{\mu \mu} \approx |U_{\mu 3}|^4 + (1 - |U_{\mu 3}|^2)^2 -  
4\left(\frac{|U_{\mu 1}|^2 }{R_{12}} + F\right) 
\left(1 - |U_{\mu 3}|^2  - F - \frac{|U_{\mu 1}|^2 }{R_{12}}\right) 
\sin^2 \frac{\Phi_{12}^m}{2}~.
\label{pmumu}
\ee
Let us underline that $F \equiv F(|U_{e1}|^2, |U_{\alpha 3}|^2)$ 
is a known function of $|U_{e1}|^2$  and $|U_{\alpha 3}|^2$ and it 
can be determined once these elements are measured.  
The contribution  of the $|U_{\mu1}|$-dependent terms to the 
probability is about 10\%. Therefore to determine $|U_{\mu 1}|^2$ 
precisely enough, the probability should be measured
with  better that 1\% accuracy.  

The correction to the formula (\ref{pmumu}) due to matter effects 
are of the order $\epsilon$.

For antineutrinos the probability is given by expression 
(\ref{pmumu}) substituting, 
$l_{12}/l_0 \rightarrow - l_{12}/l_0$, $R_{12} \rightarrow \bar{R}_{12}$, 
$\Phi_{12}^m \rightarrow \bar{\Phi}_{12}^m$ (obviously, $|U_{\alpha 
i}|=|\bar U_{\alpha i}|$). Note that in this case, above the 
resonance $(E > E^R_{12})$ we get 
\be
|\bar{U}_{\mu 1}^m|^2 \rightarrow F \approx
\frac{|U_{e3}|^2|U_{\mu 3}|^2}{1 - |U_{e3}|^2}, 
\label{jedir}
\ee  
and again the dependence on $|U_{\mu 1}|$ disappears.

In general the   aforementioned conditions 
(to measure $|U_{\mu 1}|$)  are fulfilled for the
sub-GeV atmospheric neutrinos reaching the detector through 
nadir angles between $30^o$ (for which the  baseline is tangent to the 
core) and $ 80^o$ (with $L\simeq 2000$ km). 
Indeed, for such neutrinos the phase of oscillations 
driven by $\Delta m_{sun}^2$ is of order 1: 
${\Delta m_{sun}^2 L / 2E}\sim V_e L  \sim O(1)$,  while  ${\Delta
m_{atm}^2 L/2E}\gg 1$. 
However, due to the presence of both electron and muon neutrinos 
in the initial flux, the number of observable events, 
{\it e. g.} $\mu$-like events,   depends both on survival and 
 on the conversion probabilities ($P_{e \mu}$ and $P_{\mu e}$). One 
can  easily show that  for conversion probabilities, the
 effects of interference terms, which depend on unknown 
phases, are non-negligible.
So, it is not clear,  whether atmospheric neutrino data
can help  to measure $|U_{\mu 2}|$.

\section{Do alternative methods exist?}

A straightforward (and similar to what we do in quark sector) way to
determine the elements of the MNS matrix (and
therefore the sides of the unitarity triangle) is to study the 
charged current interactions of neutrino {\it mass 
eigenstates}, $\nu_i$. Indeed, the cross-section of  the  
interaction
$$
\nu_i + X \rightarrow l  +  Y,
$$
where  $l$ is a charged lepton, is proportional to 
$|U_{li}|^2$. 
In particular, measuring  the number of electrons and muons produced 
by the  $\nu_1$-beam one can immediately find the ratio 
$|U_{e1}|/|U_{\mu 1}|$.  
To perform such a measurement one  
needs to create a beam of pure neutrino mass eigenstate 
energetic enough to produce the charged lepton, $l$. 
There are several ways to produce (in principle) a pure 
mass eigenstate beam: (i) via  adiabatic conversion, (ii) 
due to spread of the wave packets and (iii) 
as a consequence of neutrino decay.   
In general, one can also use a beam of several mass eigenstates
provided that they are {\it incoherent}. 
Processes induced by such a
beam will be determined by the moduli of matrix elements.
Effective loss of coherence  occurs due to averaging of oscillation  
of neutrinos from far objects (for which 
${\Delta  m_{sun}^2 L/ 2 E} \gg 1$). 
We will consider these possibilities  in  turn.

\subsection{Adiabatic conversion of neutrinos in matter}

In a medium with high density (larger or  much larger than the
resonance density)  mixing can be suppressed. That is, the flavor
state, produced at such a density,
coincides with the eigenstate of the instantaneous
Hamiltonian:  $\nu_f \approx \nu_{im}$.
If the density decreases slowly to zero along the path  of
neutrino, such that the adiabaticity condition is fulfilled,
the neutrino state will always coincide with the same
eigenstate:  $\nu(t) \approx \nu_{im}(t)$.
As a result, when the neutrino exits the layer (at zero density), it  will
coincide with the mass eigenstate
$\nu(t_f) \approx \nu_{im} = \nu_i $.

This happens for  solar neutrinos with energies 5 - 14 MeV
in the case of the LMA-MSW solution.  The
electron neutrinos produced in the center of the Sun are converted
to $\nu_2$-state. So, by studying the interactions of neutrinos from the 
Sun
we can measure  $|U_{e2}|$.

Obviously, usual solar neutrinos cannot produce muons.
Measurements of $|U_{\mu1}|$ and/or $|U_{\mu2}|$ will be possible,
if high   energy neutrinos ($E > m_{\mu}$)
appear in the center of the Sun and propagate adiabatically to the
surface.  Such a possibility can be realized if
massive dark matter particles, WIMPs, are  trapped inside the Sun and
annihilate  emitting  neutrinos.

Suppose that the dark matter is composed of neutralinos,
$\chi$. The neutralinos annihilate into the Standard Model
particles:   $\chi \chi \rightarrow {\rm
W}^+{\rm W}^-, {\rm Z}{\rm Z}$, $q \bar q$ etc., which in turn
decay  producing neutrinos and antineutrinos.
The energy spectra and the flavor composition
of neutrino fluxes (as well as the absolute value of the flux)
depend on the parameters of the SUSY model.
Generically,  one  expects an asymmetric  flavor composition.
Indeed, neutralinos annihilate preferably into $b \bar b$, $\tau \bar
\tau$ (and if they are massive enough also into $t \bar t$, $W^+W^-$ and 
$Z \bar Z$). Moreover, muons, pions and  kaons are absorbed or lose a  
substantial fraction of their energy  before decay. In contrast, the
$\tau$-leptons decay
before   appreciable  energy loss.  As a  result, one  expects an excess
of $\nu_{\tau}$ and approximately equal fluxes of $\nu_{e}$ and
$\nu_{\mu}$:
\be
F^0(\nu_{\tau}) = F^0(\bar{\nu}_{\tau}) >
F^0(\nu_{\mu}) = F^0(\bar{\nu}_{\mu}) \approx
F^0(\nu_{e}) = F^0(\bar{\nu}_{e}).
\label{fluxes}
\ee
At high energies ($E>100$ GeV) the  inelastic interactions inside 
the Sun are very important and 
due to differences in the cross-sections
one expects different energy spectra
for neutrinos and antineutrinos.
However, for $E_\nu < 50$ GeV  the effect of inelastic  
interaction is smaller than 10$\%$.  

Note that in contrast to the case of usual
solar neutrinos (for which pure $\nu_e$ flux is generated),
WIMPs produce a neutrino flux with a {\it complex} flavor
composition. This creates two problems:
(i) one needs to know the flavor composition which
is subject to various uncertainties;
(ii) the final flux is a mixture of mass eigenstates (and is not a pure
mass eigenstate).

Another problem  is
that only for rather low energies, the  adiabaticity 
condition is fulfilled in the resonance channel.
For  $E > 1$ GeV neutrinos cross two resonance regions inside
the sun: the  high density ($h$)-resonance associated with $\Delta
m_{atm}^2$ at density
$\rho < 30~ g/{\rm cm}^3$ and 
the low density ($l$)-resonance associated with $\Delta m_{sun}^2$ at
$\rho < 0.5 g/{\rm cm}^3$ for the LMA-MSW solution.
For definiteness we will consider the scheme with normal
mass hierarchy in which both resonances are in the neutrino channels.

The jump probability  at the resonance which characterizes the 
adiabaticity violation  can be written  
as $P_c \approx \exp(-\gamma \sin^2\theta)$, where  
\be
\label{adpar}
\gamma = 14 \left(\frac{\Delta m^2}{10^{-5} {\rm eV}^2 }\right) 
\left(\frac{1 {\rm GeV}}{E} \right)~, 
\ee
where we have used the density profile of the Sun  in \cite{john}.
(The above formula is valid only for weak violation of adiabaticity: 
$P_c\ll 1$.) For $E = 4$ GeV  in the l-resonance  associated
with $\Delta m^2_{sun} = 5 \cdot 10^{-5}$ eV$^2$ and $\tan^2 \theta=0.35$
we obtain  $P_c \approx 0.01$. 
At the  h-resonance  violation of adiabaticity is negligible, 
provided that $U_{e3}$ is not very small. We have 
$P_{c} \sim 10^{-7}$, for $|U_{e3}|^2 = 0.03$.  
  
So, the adiabaticity
violation effects are below 5 \% for $E <   5$ GeV.
In the antineutrino (non-resonant)  channel
the adiabaticity violation occurs at  larger energies.

The effects of adiabaticity violation lead to the appearance
of interference terms which depend on unknown complex phases.
Therefore one needs to select low energy events. On the other hand,
even for light neutralinos, $m_\chi\sim 50$ GeV,
we expect that only about 30\% (or less) of neutrinos have energies less 
than 5  GeV. This diminishes  statistics
and hence the efficiency of the method.
Notice also that large underwater (ice) detectors have rather
 high energy thresholds, so  detection of  GeV neutrinos
is may be  problematic.

In what follows we present a simplified consideration (neglecting  
the inelastic interactions)
to illustrate  possibilities of the method and its shortcomings.
A detailed analysis will be given elsewhere \cite{yas}.

Using  relations  (\ref{fluxes}) we can write the $\nu_{\beta}$ flux
($\beta = e, \mu, \tau$) at the Earth as
\be
F_{\beta} = \sum_{\alpha} P_{\beta \alpha}
F^0_{\alpha}  =
F^0 +  \Delta F^0 P_{\beta \tau},
\label{betatau}
\ee
where $F^0 = F^0_{\mu} = F^0_e $ is the common flux of 
the electron and muon   neutrinos,
$\Delta F^0 \equiv F^0_{\tau} -  F^0_{\mu}$ and
$P_{\beta \alpha}$ is the
$\nu_{\alpha}   \rightarrow \nu_{\beta}$
conversion probability on the way from the production region in the
center of the Sun to the detector on the Earth.
(Here for simplicity we do not consider the Earth matter effect.)
A similar expression can be written for the antineutrino  channels.

Neutrinos from  WIMP annihilation can be detected by
large underwater and ice  detectors via
 charged current interactions. In these detectors 
the rates of $\mu$-like   events will be measured.
The detectors will not be able to
identify the charge of the produced lepton and therefore
we  need to sum  the  neutrino and antineutrino fluxes in our analysis.
Using Eq. (\ref{betatau}) we can write the expression for the rate
of  the $\mu$-like events as
\be
N_{\mu} = N_{\mu}^0 +
\int d\Omega \Delta F^0 (P_{\tau \mu} \sigma_{\mu} +
P_{\bar\tau \bar\mu} \bar{\sigma}_{\mu}),
\label{numb}
\ee
where
\be
N_{\mu}^0 =
\int d\Omega F^0 (\sigma_{\mu} + \bar{\sigma}_{\mu})
\label{numb_0}
\ee
is the rate without oscillations. In the above equations, $\sigma_{\mu}$ 
and $\bar{\sigma}_{\mu}$  are the cross-sections of the charged current
interactions of neutrinos and antineutrinos, respectively.
Here $\int d\Omega$ includes the integration over the neutrino
energy, the angle between the neutrino and the produced muon and  the
energy  of muon.
One should also include the efficiency of
detection and  the energy resolution function.

Let us  find the transition probability, $P_{\beta \tau}$,
which determines  according to (\ref{betatau}) the oscillation effects. 
The general expressions for  the probabilities $P_{\beta \alpha}$
are  given in  Ref.~\cite{gouvea}.
Here for illustrative purposes we will consider
the  case of pure adiabatic propagation  in the Sun.

For  $E < 5$ GeV  and the relevant $\Delta m^2$ 
 the oscillatory terms will be averaged out, in
particular, due to finite energy resolution of the detector and  change
of the distance between the Sun and the Earth. 
Taking into account this averaging effect we find
the $\nu_{\tau} \rightarrow \nu_{\beta}$  conversion
probability in the adiabatic limit: 
\be
P_{\beta \tau} = \sum_{i = 1,2,3} |U_{\beta i}|^2
|U_{\tau i}^m|^2,
\label{prob}
\ee
where $U_{\tau i}^m$ is the mixing parameter in matter
at the production region.

The  density
in the production region  is much higher than  the resonance densities,
so that  the mixing  is strongly suppressed.
Considering the level crossing diagram, it is easy to show that
$\nu_{3m} \approx - \nu_e$,
$\nu_{2m} \approx \nu_{\tau}'$ and $\nu_{1m} \approx -\nu_{\mu}'$.
{}From these relations and the definition of $\nu_{\mu}'$ and
$\nu_{\tau}'$ (\ref{basis}) we obtain 
\be
|U_{\tau 1}^m| = \frac{|U_{\mu 3}|}{\sqrt{1 - |U_{e3}|^2}},~~~~~
|U_{\tau 2}^m| = \frac{|U_{\tau 3}|}{\sqrt{1 - |U_{e3}|^2}}, ~~~~~
U_{\tau 3}^m \approx 0.
\label{mixing0}
\ee
Inserting (\ref{mixing0}) into (\ref{prob}) we find the
probability of the $\nu_{\tau} \rightarrow \nu_{\mu}$
conversion:
\be
P_{\tau \mu} =
\frac{1}{1 - |U_{e3}|^2} \left[|U_{\mu 3}|^2 (1 - |U_{\mu
3}|^2) + (1 - 2|U_{\mu 3}|^2) |U_{\mu 2}|^2\right]
\label{adiab_mu_tau}~.
\ee
Since the atmospheric mixing is close to maximal:
$|U_{\mu 3}|^2 \sim 1/2$,  the dependence of the probability
$P_{\tau \mu}$ on $|U_{\mu 2}|^2$  is weak.

In the antineutrino channel, at high densities we have
$\bar{\nu}_{1m}  \approx  \bar{\nu}_e$,
$\bar{\nu}_{2m} \approx \nu_{\mu}'$ and
$\bar{\nu}_{3m} \approx \bar{\nu}_{\tau}'$.
Consequently, the mixing elements equal:
\be
|\bar{U}_{\tau 1}^m| \approx 0 ,~~~~~
|\bar{U}_{\tau 2}^m| \approx
\frac{|U_{\mu 3}|}{\sqrt{1 - |U_{e3}|^2}} ,~~~~~
|\bar{U}_{\tau 3}^m| \approx
\frac{|U_{\tau 3}|}{\sqrt{1 - |U_{e3}|^2}} .
\label{mixing1}
\ee
Using (\ref{prob}) we find the probability
of the $\bar{\nu}_{\tau} \rightarrow \bar{\nu}_{\mu}$
conversion:
\be
\bar{P}_{\tau \mu} =
\frac{|U_{\mu 3}|^2}{1 - |U_{e3}|^2}
\left[1 - |U_{\mu 3}|^2 - |U_{e3}|^2 +  |U_{\mu 2}|^2 \right].
\label{aadiab_mu_tau}
\ee
Here  $|U_{\mu 2}|^2$ appears with a relatively large coefficient.

In the adiabatic limit,  the conversion probabilities do not
depend on energy and the expressions for the rate  of events can be
written as
$$
N_{\mu} \approx  N_{\mu}^0 +
\frac{|U_{\mu 3}|^2 (1 - |U_{\mu 3}|^2)}{1 - |U_{e3}|^2}
\int d\Omega \Delta F^0
(\sigma_{\mu} +  \bar{\sigma}_{\mu}) +
$$     
\be
\frac{|U_{\mu 2}|^2}{1 - |U_{e3}|^2} \int d\Omega
\Delta F^0 [(1 - 2|U_{\mu 3}|^2)
\sigma_{\mu} + |U_{\mu 3}|^2 \bar{\sigma}_{\mu}].~~~~~~
\label{numb-ad}
\ee
According to (\ref{numb-ad}), the relative effect of the term proportional 
to $|U_{\mu 2}|^2$
is  suppressed by smaller value of the antineutrino
cross-section   $\bar{\sigma}_{\mu}/\sigma_{\mu} \sim 1/2$.

The relative contribution to number of events from  the term
which depends on $|U_{\mu 2}|^2$ at low energies (see (\ref{numb-ad}))
is 
\be
r \approx   \frac{|U_{\mu 2}|^2 |U_{\mu 3}|^2}{N_{\mu}^0}
\int d\Omega \Delta F^0 \bar{\sigma}_{\mu}.
\label{number-ad}
\ee      
For larger energies $E > 7$ GeV, in the neutrino channel the  effects of 
the adiabaticity violation
are $\stackrel {>} {\sim} 10\%$, {\it i. e.} larger than the level of 
required  accuracy in the 
determination of the mixing elements. In the antineutrino channel
the adiabaticity violation is weaker. So if  detector is able to
identify the charge of lepton, and consequently, to select
antineutrino events, one will be able to perform
better measurements. In particular, events with higher energies can be
studied.


Let us comment on the possibility to detect neutrinos from WIMP
annihilation
and to measure $|U_{\mu2}|^2$.
The event rates due to these
neutrinos in a ${\rm km}^3$-size detector can be as large as few $10^3$
events/year \cite{gondolo}, although the rate is very model dependent.    
If $\Delta F^0/F^0 \sim 0.2 - 0.5$, the contribution of the term
sensitive to $|U_{\mu2}|^2$ is about 10 \%.
Therefore  $|U_{\mu2}|^2$ can be determined with accuracy
10\% at best, provided that all other involved parameters  are known. In
particular, one should know
the  original flux  $F^0_{\mu}$, and the difference of fluxes
$F^0_{\tau} -  F^0_{\mu}$ as functions of energy.

There are several possible ways to deduce information
about the ratio of original fluxes ${\Delta F^0/ F^0}$:

1). Theoretical predictions:
In principle, future high energy experiments at colliders 
({\it e.g.} LHC),
as well as results of the direct searches for  dark matter particles
will help   to measure the mass and the composition (Higgsino-like
{\it versus} gaugino-like) of  neutralinos.
This will allow  to predict the {\it relative}  neutrino 
fluxes from
annihilation.
  
2).  Information on {\it relative} neutrino fluxes from WIMP
annihilation can be obtained by detecting  neutrinos
from WIMP annihilation in the Earth center.

These studies cannot  determine the  absolute value of the original flux 
($F^0$). Once we obtained by the aforementioned methods the value of 
${\Delta  F^0/ F^0}$,  
we can try to  measure both  $|U_{\mu 1}|^2$
and the original fluxes from studies of solar neutrinos themselves.
If the detector is able to identify the flavor \cite{wang},
we can compare the rates  of $\mu$-like with
$\tau$-like or $e$-like events to find the total flux and the value of  
$|U_{\mu 1}|^2$.

\subsection{Spread of the wave packets} 

Bunches of  neutrino mass eigenstates can be 
obtained as a result of difference in the  group velocities. 
Neutrinos with a mass squared difference 
$\Delta m^2$ but the same energy $E$, produced in a source at the same
time, will arrive at the detector with a time difference 
\be 
\label{spread}  
\Delta t = 0.1 {\rm sec} \left({L\over 10^{28} \ \ {\rm cm} 
}\right)\left({\Delta m^2  
\over 3 \times 10^{-3} \ \ {\rm eV}^2}\right)
\left({100\ \ {\rm MeV} \over E}\right)^2 \ .
\ee
Here $L$ is the distance from the source.  
If the time during which neutrinos are produced at 
the source, $\tau_p$, is  considerably  smaller  
than $\Delta t$  ($\tau_p <\Delta t$),  
and if the energy spread is 
small enough (or the detector is able to select 
neutrinos of certain energy), 
neutrinos  will arrive at the detector in bunches: the heavier
neutrinos  arrive after the light ones.  
We can measure the numbers of charged leptons produced by different 
bunches via the charged current interactions. 
Thus, the  ratio of number of muons and $\tau$-leptons   
 produced by the first bunch 
gives   $|U_{\mu 1}/ U_{\tau 1}|$.  
Similarly, the second and  third bunches 
give information 
about $|U_{l2}|$ and $|U_{l3}|$,  respectively. 
The number of charged   leptons of a given flavor, $l$, produced by  
the first  and  second bunches  is proportional 
to   $|U_{l1}|/|U_{l2}|$, etc.

According to (\ref{spread}), the time difference in arrival of 
the bunches for  $\Delta m^2 = 3 \times 10^{-3}$ ${\rm eV^2}$, 
$E=100$ MeV and $L = 10^{28}$ cm equals  $\Delta t$ = 0.1 sec.  
So, the duration of the neutrino pulse  
should be smaller than 0.1 second. 
Moreover,  the number of  events induced by a single  pulse 
should be large enough. 
It is not clear if the required  sources of neutrinos exist.

\subsection{The decay of neutrinos} 

The neutrino decay  provides another possibility 
to  get pure   beam  of   mass eigenstates.	
In the minimal extension of the Standard Model (in which  neutrinos are 
massive and there are right-handed 
 neutrinos) neutrinos can decay  radiatively:  $\nu_j \rightarrow \nu_i + 
\gamma$. However,  the life time  is extremely large:  
$\tau_\nu >  10^{45}$ s.  
In certain  extensions of the SM the radiative decay 
or $3\nu$-decay may be much faster, however, according to astrophysical 
bounds, the lifetime of radiative decay must be much larger than 
the age of the Universe (see review \cite{roulet}). 

The  decay which  satisfies all the bounds and is relevant for 
our analysis  is the Majoron decay~\cite{majoron,pak}: 
\be 
\nu_i \rightarrow \bar{\nu}_j +J,
\ee
where $\nu_i$ and $\nu_j$ are mass  eigenstates, and $J$ 
is the Majoron.  

Let us assume that $\tau_\nu \gg 10^{-3}$ sec, so 
the  neutrinos from the Sun do not decay and the solar  
and atmospheric anomalies are explained by 
oscillations while neutrinos from very far 
sources ({\it i.e.}, the  gamma-ray 
bursters,  the Active Galactic Nuclei and supernovae) decay 
before reaching the Earth.  Then  at the detectors the neutrino flux 
from the far source is 
composed   only of the lightest neutrinos 
$\nu_1$ and $\bar{\nu}_1$. 

The $\gamma$-ray bursts may be  accompanied by a 
flux  of energetic neutrinos \cite{proceed,waxman}. 
Taking  the  distance of the $\gamma$-ray burster from the Earth
to be of order  $10^{28}$ cm, one finds that all heavy neutrinos 
will decay if the lifetime of neutrino at  rest, $\tau_\nu$,  
satisfies the inequality 
\be 
\tau_\nu \stackrel {<} {\sim} 10^{16} {\rm sec} ~ {m_\nu \over E}. 
\label{ineq}
\ee 
Let us evaluate this bound both for hierarchical and quasi-degenerate    
neutrino mass spectrum setting $E \sim 1$ TeV. In the hierarchical case  
$m_1\simeq 0$, $m_3 \sim 0.05$ eV and  $m_2 \sim 0.007$ eV so 
from (\ref{ineq}) we find that in order to let $\nu_3$ decay   
 $\tau_3$ should be $ \stackrel {<}{\sim}  10^{2}$ sec,  while for 
$\nu_2$, the bound   is  $\tau_2 <  10$ sec.
For quasi-degenerate spectrum with $m_1\simeq m_2\simeq m_3=1$ eV, the 
bound is weaker: $10^4$ sec.

It was estimated~\cite{waxman}  that the flux of  
neutrinos with the TeV scale energies   
from an individual gamma burster at cosmological distance $z \sim 1$ 
produces $10^{-1}-10$ muons in 1 ${\rm km}^3$-size 
detectors. Since these neutrinos are correlated in  
time with the $\gamma$-ray bursts and  aim at 
the same source, they can be distinguished from 
neutrinos produced by other  sources. The rate of 
$\gamma$-ray bursts detectable on the Earth is 
$\sim 10^3 /{\rm year}$ so the statistics are  
fairly high and we can deduce results based on 
studies of  such neutrinos.

The large scale  detectors cannot identify 
the  charge of produced leptons, so in practice  $\nu_1$ and 
$\stackrel {-} \nu_1$ signals will be summed up.
Unfortunately, with present design for 1 ${\rm km}^3$-size detectors, it
is
hardly possible to identify the flavor of the detected neutrinos, for the 
energy range which we are
interested. However, there are methods which open a possibility  to build
a large detector with flavor  identification \cite{wang}.

Let  us assume that these technical problems will  be solved and that  
future detectors will be able to discriminate between different flavors.
In the presence of the decay  
 only the flux of the lightest neutrino, $\nu_1$, will arrive at the
Earth. Then the ratio of $\mu$-like events to 
$\tau$-like events produced by this flux equals  
\be 
\label{rat} 
{\mu{\rm -}{\rm like \ \ events} 
\over 
\tau{\rm -}{\rm like \  \ events}} = 
{|U_{\mu 1}|^2 \over |U_{\tau 1}|^2}, 
\ee
where we have taken into account  that for 
high energies,  neutrinos of different flavors have 
nearly equal cross-sections:
$\sigma (\nu_\mu)\simeq \sigma(\nu_\tau)$ and $\sigma 
(\stackrel {-} {\nu_\mu})\simeq \sigma (\stackrel {-} {\nu_\tau})$.

Thus,  if the detectors are able to identify  $\tau$-like events, 
we will be able to  measure the ratio $|U_{\mu 1}/U_{\tau 1}|$. 
Using this ratio,  the unitarity condition 
$|U_{e 1}|^2+|U_{\mu 1}|^2+|U_{\tau 1}|^2=1$,
and $|U_{e  1}|^2$ determined  by KamLAND, 
one   can derive the value of $|U_{\mu 1}|$.  
Then   $|U_{\mu 2}|^2 = 1 - |U_{\mu 1}|^2 - |U_{\mu 3}|^2$.

Similarly, for the ratio of $\mu-$like to $e-$like events we have
\be
\label{mouse}
{\mu{\rm -}{\rm like \ \ events}
\over e{\rm -}{\rm like \  \ events}} =
{|U_{\mu 1}|^2 \over |U_{e 1}|^2},
\ee
where we have used $\sigma (\nu_\mu)\simeq \sigma (\nu_e)$ and
$\sigma (\bar \nu_\mu)\simeq \sigma (\bar \nu_e)$. 
This ratio can be used to determine $|U_{\mu 1}|^2$ immediately, 
once $|U_{e 1}|^2$ is known. Unfortunately, identification of 
$e$-like events is very difficult.

Let us emphasize  that the analysis based on (\ref{rat}) and (\ref{mouse}) 
does not depend on astrophysical details (neutrino production mechanism, 
etc.). 
However, one should make sure that all heavy neutrinos have decayed 
on their way to  the Earth. A  check can be based on 
ratio of fluxes (eventually numbers of events). 
If neutrinos are stable we expect 
$F(\nu_e):F(\nu_\mu):F(\nu_\tau) \simeq 1:1:1$ \cite{athar},  
while in the case of decay 
$F(\nu_e):F(\nu_\mu):F(\nu_\tau)=|U_{e1}|^2: |U_{\mu 1}|^2:|U_{\tau 1}|^2  
\simeq 1:{1 \over 2}:{1 \over 2}$. 
The above analysis was based on  assumption that 
neutrinos from all  $\gamma$-ray bursters  
decay before reaching the Earth. It may happen,  
however,  that due to spatial distribution  of  
sources,  the degree of decay can be different for  different 
sources.  From  a single burst only 
few neutrinos can be detected,  
so studying neutrinos associated with only one $\gamma$-burst event,  it
is impossible to establish 
the existence  of the decay. 
This can be done on the basis of observations of many bursts.  
The sources  can be divided into two groups: close sources 
and far sources
(the distance of the source can be measured by its redshift).  
Studying  the flavor composition of neutrino fluxes from 
these   two groups, one  can check the stability of 
neutrinos. 
There are other  measurements which can  shed  light on 
the decay  rate of  neutrinos \cite{pak,pas}.

Note that  this analysis  does not 
depend on the solution of the solar problem.

\subsection{Loss of coherence; averaged oscillations}

Let us consider  {\it stable} or meta-stable neutrinos 
produced by  cosmological sources. For example, consider again  the 
neutrinos with $E\sim 1$ TeV accompanying the $\gamma$-ray
bursts \cite{waxman}.
  For such neutrinos the oscillation length is much smaller than 
the distance from  the source, $L\sim 10^{28}$  cm  
(even for the VAC solution of the solar neutrino problem, 
$\Delta m_{sun}^2 L/E \gg 1$). 
Consequently,  all the oscillatory terms in the probabilities 
will be averaged out.   
Furthermore, according to existing models of the bursters, 
the neutrinos  are produced in the envelope of the star with density  
$\rho \sim 10^{-7}$ g ${\rm cm}^{-3}$ and radius 
$\sim 10^{13}$ cm and therefore the matter effects inside the source  
are negligible~\cite{cecilia}. As a consequence, the oscillation 
probabilities for neutrinos 
$(\nu_{\alpha} \rightarrow \nu_{\beta})$ and 
antineutrinos $(\bar{\nu}_{\alpha} \rightarrow \bar{\nu}_{\beta})$ 
take the form   
\be
\label{easily}
P_{\alpha \beta}  = \bar{P}_{\alpha \beta} = 
\sum_{i} |U_{\alpha i}|^2|U_{\beta i}|^2.
\ee
(Here we used $|\bar U_{\alpha i}|=|U_{\alpha i}|$.)  
In particular,  
\be
\label{mu-mu}
P_{\mu \mu} = \sum_{i} |U_{\mu i}|^4 = 
K_{\mu\mu} - 2 |U_{\mu2}|^2 |U_{\mu 1}|^2~ 
\ee  
and 
\be
\label{e-mu}
P_{e\mu} = \sum_{i} |U_{\mu i}|^2|U_{e i}|^2  = K_{e\mu} - 
 |U_{\mu2}|^2 (|U_{e 1}|^2 - |U_{e 2}|^2),  
\ee 
where  $K_{\mu\mu}$  and  $K_{e\mu}$ are known functions of 
 $|U_{e1}|$, $|U_{e2}|$, $|U_{e3}|$, $|U_{\mu 3}|$ and do not depend 
on $|U_{\mu1}|^2$ and  $|U_{\mu2}|^2$. The probability 
$P_{ee}$ does not depend on $|U_{\mu1}|^2$ and  $|U_{\mu2}|^2$. 

The probabilities (\ref{easily}),  (\ref{mu-mu}), 
(\ref{e-mu}) have the following  properties which play a  key role 
in our calculations:  

(i) $P_{\alpha \beta} = P_{\beta \alpha}$;

(ii) probabilities for neutrinos and antineutrinos are equal; 

(iii) the probabilities do not depend on energy.

Let us calculate the number of charged current events produced by  
neutrinos from  $\gamma$-ray bursters in the detectors. 
We assume that the source  produces (differential) fluxes of 
electron neutrinos, $F^0_e$, and antineutrinos, $\bar{F}^0_{e}$, 
as well as muon neutrinos, $F^0_{\mu}$, and antineutrinos,  
$\bar{F}^0_{\mu}$, whereas the fluxes of $\tau$-neutrinos and
$\tau$-antineutrinos 
are negligible. Using the properties of the oscillation probabilities 
listed above and summing up neutrino and antineutrino contributions  
we can write for the number of $\mu$-like events 
\be
(\mu{\rm -like \ \ events}) = 
P_{\mu \mu} \left(\int F^0_{\mu} \sigma dE + 
\int \bar{F}^0_{\mu} \bar\sigma dE\right) + 
P_{\mu e} \left( \int F^0_{e} \sigma dE +  
\int \bar{F}^0_{e} \bar\sigma dE \right)~,
\ee
where $ \sigma =  \sigma (\nu_e) \simeq \sigma (\nu_\mu)
\simeq \sigma(\nu_\tau)$ and $\bar\sigma =   
\sigma (\bar\nu_e)\simeq\sigma (\bar\nu_\mu )\simeq 
\sigma(\bar\nu_\tau)$ are the cross-sections for neutrinos and
antineutrinos. Similar  expressions can be written 
for the number of $e$-like and $\tau$-like events.  
Notice that the oscillation probabilities factorize out of the  
integrals over energy.

For the ratios of event numbers we can write 
\be 
{\mu{\rm -like \ \ events }\over e {\rm -like \ \ events }}= 
{P_{\mu \mu}A+P_{ \mu e}  \over 
P_{ee}+P_{e\mu} A } \label{tachin}\ee
and 
\be
{\tau{\rm -like \ \ events }\over e {\rm -like \ \ events }}= 
{P_{\tau \mu}A+P_{\tau e}  
\over P_{ee}+P_{e \mu} A }~, 
\label{shabih}
\ee
where 
$$ 
A \equiv 
\frac{\int F^0_\mu \sigma dE + \int  F^0_{\bar\mu} 
\bar{\sigma} dE}{\int F^0_e \sigma dE + \int F^0_{\bar e} \bar\sigma dE}
$$ 
and the probabilities are defined in (\ref{easily}). The ratios in 
(\ref{tachin}) and  (\ref{shabih}) 
are functions of $A$ and $|U_{\mu2}|$. Presumably 
all other mixing parameters will be measured by terrestrial experiments 
described in sect. 3.1 - 3.4. 
The astrophysical information (and uncertainties)  is contained in $A$
and it will be probably difficult (if possible) to 
predict this quantity. So basically we should  consider $A$ as an unknown
parameter. 
If future  detectors are able to identify flavor \cite{wang}, 
we can  determine  two ratios (\ref{tachin}) and  (\ref{shabih})  
and  $A$ and  $|U_{\mu2}|$.  
Additional cross checks of results can be done 
if the detector is  able to identify the charge of the produced lepton.

Note that this method works for all solutions of the solar 
neutrino problem.

\section{Discussions and conclusions}

Reconstruction of the unitarity triangle is the way to 
establish CP-violation alternative to the one based on the 
direct measurements of the CP- or T- asymmetries.  

Properties of the leptonic unitarity triangles have been studied.  
Our estimates show  that for maximal allowed value of  $|U_{e3}|$ 
and maximal CP violation ($\delta_D = 90^{o}$) 
a precision better than  10\% in measurements of the sides 
of the triangle will allow us to  establish  CP-violation  
at the 3$\sigma$ level. 

We have considered the possibility to reconstruct the triangle in future 
oscillation experiments.  
For this in the three neutrino context, one needs to measure the absolute
values of the mixing matrix elements:   
$|U_{e2}|$, $|U_{e3}|$, $|U_{\mu 2}|$, $|U_{\mu 3}|$.    
The elements of the first side can be obtained from normalization. 
In general, the oscillation probabilities 
depend both on these absolute values  and 
on the unknown relative phases. We  suggest   
configurations of experiments: channels of neutrino oscillations, 
neutrino energies, baselines, and averaging conditions, for which  
corrections to the probabilities  which depend on  
unknown phases are sufficiently small.  
We estimate that for value of $|U_{e3}|$ at the present upper bound and  
$\delta_D = 90^{o}$, the 
elements $|U_{e2}|$, $|U_{e3}|$, $|U_{\mu 2}|$, $|U_{\mu 3}|$ 
should be measured with better than 
3 - 5 \% accuracy to establish CP-violation.

We have found that 

1). The third side of the triangle,  $|U_{e3}^* U_{\mu 3}|$,  can be
directly measured in $\nu_{\mu} - \nu_e$ 
oscillations 
driven by $\Delta m^2_{atm}$. In this case, the relative corrections to 
the probability 
which depend on the unknown phase, $\Delta P_{\delta}/P$, are as large as 
$|U_{e3}|^{-1}\epsilon \sim 10\%$. 
This substantially restricts the application of the method.  
Relative corrections could be suppressed at high energies: 
$E > E^R_{13}$ by an additional factor $E^R_{13}/E$.  
An alternative way to determine $|U_{e3}^* U_{\mu 3}|$ would be
independent measurements of   
$|U_{e3}|$ and  $|U_{\mu 3}|$ in different experiments.

2). The element  $|U_{e3}|$ can be measured in 
studies of the ($\nu_e \rightarrow \nu_e$) 
survival probability 
at low energy (reactor) experiments. The uncontrolled corrections 
to the probability are of  order of $\epsilon$.

3). $|U_{\mu 3}|$ can be determined by measurements of the 
$\nu_{\mu} - \nu_{\mu}$ survival probability in the $\nu_{\mu}$ oscillations 
driven by $\Delta m_{atm}^2$. Both in the low energy 
$(E < 500$ MeV)
and in the high energy $(E >  5$ GeV) regimes
the relative uncontrolled corrections to the probability are of the order
of
$\epsilon$.

4). The ratio of elements $|U_{e1}|$  and $|U_{e2}|$ can be measured 
rather precisely by  KamLAND and  by
further solar neutrino studies. 
Then $|U_{e1}|$  and $|U_{e2}|$  can be 
obtained using  the normalization condition and  the
value of $|U_{e3}|$ (measured in other experiments).  

5). The determination of $|U_{\mu 1}|$ and $|U_{\mu 2}|$ is the most
difficult 
part of the program.   These quantities could be measured studying  the 
$\nu_{\mu}$  disappearance  at   low
energies  ($E < 500$ MeV) in  experiments with a base-line $L > 2000$ km.
The uncontrolled corrections, here, are relatively small
($< \epsilon$). 

The reconstruction of the unitarity triangle requires a series of
measurements which differ from direct measurements of CP- and
T-asymmetries. Indeed,  

\begin{itemize}

\item
information on the absolute values of matrix elements follows mainly
from studies of the {\it survival} probabilities; 

\item
both neutrino and antineutrino beams 
give the similar results, so that one can work with a neutrino beam or an 
antineutrino beam  or with some combination of them;   

\item
averaging of the oscillatory terms and the loss of coherence  help for 
determination of the relevant parameters. 

\item 
the method works better 
(in a sense that uncontrolled corrections are smaller)
for smaller value of 
ratio  $\epsilon = \Delta m_{sun}^2/\Delta m_{atm}^2$ which 
determines relative uncontrolled corrections.

\end{itemize}

These factors inhibit direct observations of CP-violation. 
In this sense, the method of the unitarity triangle 
is complimentary to the direct determination of CP-violation 
from measurements of asymmetries.

Clearly, the proof of feasibility of the method requires further studies,  
and especially, calculations of the effects in specific experiments. 
In any case, realization of the program will not be easy.

The  ideal direct  way to measure the  
absolute values of matrix elements  would be 
experiments with beams of pure mass eigenstates of neutrinos. 
Charged current  interactions of these beams will help to determine 
immediately 
the value of $|U_{\alpha i}|^2$ (or the ratios $|U_{\alpha 
i}|^2/|U_{\beta i}|^2$ if the 
absolute value of the neutrino flux is not known). 
However the masses of neutrinos are extremely small (much smaller 
than  the quark case), and therefore neutrinos are produced and propagate
in the flavor states, {\it i.e.,} the coherent states of mass eigenstates. 
Separation of the mass eigenstates is a non-trivial problem. 
We have made some suggestions  based on  spread of the neutrino wave
packets, neutrino decay, adiabatic conversion and  loss of the coherence 
due to averaging of oscillations. 

In this connection  we have considered several possibilities related  
to high energy extra-terrestrial neutrinos: 

1). Neutrinos from possible annihilation of WIMP's in the center of the sun 
or the earth.

2). Decaying neutrinos from cosmological sources. 

The oscillation probabilities  for stable neutrinos from far sources (for
which $\Delta  m_{sun}^2 L/2 E \gg 1$) dependent only on the 
absolute  
values of the mixing  matrix elements. We have considered the neutrinos 
accompanying gamma-ray bursts, as an example.

These possibilities require large ($\sim 1 {\rm km}^3$-size) 
underwater (ice) detectors which enable us 
to identify  the flavor of the interacting neutrino.

\section*{Acknowledgement}
The authors are grateful to S. P. Mikheyev for valuable comments. 
One of us  (Y. F.) would like to thank M. Peskin and G. Gratta for
useful discussions. We would like to thank F. Petriello 
for careful  reading the manuscript.

\end{document}